\documentclass[11pt, a4paper]{article}
\usepackage{jheppub}
\usepackage{braket}

\begin{document}

\title{Cutoff Scale of Quadratic Gravity from Quantum Focusing Conjecture}

\author[a,b]{Takamasa Kanai,}
\emailAdd{kanai.takamasa.l9@sic.shibaura-it.ac.jp}
\affiliation[a]{Department of Mathematics, Nagoya University, Nagoya 464-8602, Japan}

\author[b]{Kengo Maeda,}
\emailAdd{maeda302@sic.shibaura-it.ac.jp}
\affiliation[b]{Faculty of Engineering, Shibaura Institute of Technology, Saitama 330-8570, Japan}

\author[c]{Toshifumi Noumi}
\emailAdd{tnoumi@g.acc.u-tokyo.ac.jp}
\affiliation[c]{Graduate School of Arts and Sciences, University of Tokyo, Komaba, Meguro-ku, Tokyo 153-8902, Japan}

\author[a]{and Daisuke Yoshida}
\emailAdd{dyoshida@math.nagoya-u.ac.jp}

\preprint{UT-Komaba/24-3}

\abstract{
We derive the cutoff length scale of the quadratic gravity in $d \geq 5$ dimensional spacetime by demanding that the quantum focusing conjecture for the smeared quantum expansion holds at the classical level.
The cutoff scale has different dependence on the spacetime dimension depending on the sign of the coupling constant of the quadratic gravity. We also investigate a concrete example of the 5-dimensional Schwarzschild spacetime and directly confirm that the quantum focusing conjecture holds when the quantum expansion is smeared over the scale larger than our cutoff scale.
}
\maketitle

\section{Introduction}
The focusing of the null geodesics plays an important role to understand the various properties of the spacetime in General Relativity, such as the singularity theorem by Penrose~\cite{Penrose:1964wq} and the area law of the black hole~\cite{Hawking:1971tu, Hawking:1971vc}. See also, e.g., the text books \cite{Hawking:1973uf, Wald:1984rg} for the review on these topics. 
The focusing of the null geodesics is followed from the null energy condition. However, the null energy condition is violated by considering the quantum effects, for example by the Casimir effect \cite{Casimir:1948dh} and the Hawking radiation \cite{Hawking:1974rv,Hawking:1975vcx}. Thus, to understand the general property of the spacetime in the semi-classical regime, it is important to figure out an extension of the null energy condition, or the focusing property of the null geodesics, which is applicable there.

It is known that the area of the event horizon of the black hole, up to the constant factor, can be understood as the entropy, which is called the Bekenstein--Hawking entropy~\cite{Bekenstein:1972tm,Bekenstein:1973ur,Bekenstein:1974ax,Hawking:1974rv,Hawking:1975vcx}. In this view, the area law of the black hole event horizon can be understood as the second law of the black hole thermodynamics \cite{Bekenstein:1972tm,Bardeen:1973gs,Bekenstein:1973ur,Bekenstein:1974ax}.
However, as is mentioned above, the area law, and hence the second law of the black hole thermodynamics, will be violated by the Hawking radiation \cite{Hawking:1974rv,Hawking:1975vcx}.
Bekenstein proposed the generalized second law \cite{Bekenstein:1972tm,Bekenstein:1973ur,Bekenstein:1974ax}, which states that the generalized entropy, that is the entropy for the total system including both the black hole and matter, will not decrease in any physical process.
When the black hole evaporates with the Hawking radiation, the second law of the black hole thermodynamics itself is not satisfied. However, the entropy of the total system is expected to be non-decreasing because of the increase of the entropy of the Hawking radiation emitted from the black hole. Thus, the generalized second law can be understood as a generalization of the area law in the classical general relativity to the semi-classical regime.

A lesson from the generalized second law is that a geometrical property of the area in the classical general relativity can be promoted to a property of the generalized entropy in the semi-classical regime. This idea leads to the quantum focusing conjecture \cite{Bousso:2015mna}. The quantum focusing conjecture states that the quantum expansion, defined based on the generalized entropy, is not increasing along the null geodesics.
Here we note that the generalized entropy is formulated not only for the event horizon of the black hole but also the general surface for an arbitrary spacetime. The quantum focusing conjecture can be regarded as an extension of the converging property of null geodesics in the classical general relativity, that is, the property that the expansion of the null geodesics, defined as the changing rate of the infinitesimal cross sectional area of the null geodesics, is not increasing under the null energy condition. We will review the quantum focusing conjecture in the section \ref{QFC}.
Note that the quantum focusing conjecture can be translated into the condition for the energy momentum tensor of the quantum matter at a specific point. This condition is called the quantum null energy condition~\cite{Bousso:2015mna,Bousso:2015wca,Koeller:2015qmn, Fu:2017evt, Balakrishnan:2017bjg} and regarded as an extension of the null energy condition in the semi-classical regime.

The generalized entropy is composed of the gravitational entropy and the entanglement entropy of the quantum fields. In this paper, we focus on the classical limit, $\hbar \rightarrow 0$, and hence the contribution from the quantum entanglement entropy is omitted. Even in the classical limit, there is a major difference between the area and the generalized entropy because the gravitational entropy depends on the gravitational action itself.
Thus, the generalized entropy is sensitive to the higher derivative corrections to the Einstein--Hilbert action.
In the case of the black hole event horizon, the gravitational entropy with respect to the covariant gravitational action can be calculated by Wald's Noether charge approach \cite{Wald:1993nt, Iyer:1994ys, Iyer:1995kg} or the field redefinition method \cite{Jacobson:1993vj}. 
On the one hand, in the development of AdS/CFT correspondence \cite{Maldacena:1997re,Gubser:1998bc,Witten:1998qj}, the generalized entropy with respect to the Ryu-Takayanagi surface, or its generalization, is characterized as the holographic dual of the entanglement entropy of the boundary CFT, and hence, it is called holographic entanglement entropy \cite{Ryu:2006ef, Fursaev:2006ih, Headrick:2010zt, Lewkowycz:2013nqa}. The effect of the higher derivative corrections to the holographic entanglement entropy is studied in Refs.~\cite{Hung:2011xb,Fursaev:2013fta, Bhattacharyya:2013gra, Alishahiha:2013zta, Dong:2013qoa, Camps:2013zua, Bhattacharyya:2013jma,Bhattacharyya:2014yga, Miao:2014nxa, Dong:2015zba}. In particular, the formula proposed by Dong \cite{Dong:2013qoa} and Camps \cite{Camps:2013zua} enables us to calculate the holographic entanglement entropy with respect to the gravitational action expressed by the Riemann tensors. See Ref.~\cite{Miao:2014nxa} for the entropy formula for more general gravitational action. By interpolating the Dong--Camps entropy formula to the stationary black hole event horizon, one can reproduce Wald's entropy formula. Hence, in this paper, we adopt the Dong--Camps formula as the gravitational entropy for the arbitrary surface. Note that the entropy formula for dynamical black holes is discussed in Ref.~\cite{Hollands:2024vbe}.

Considering the gravitational entropy with respect to the Gauss-Bonnet gravity, it was shown in Ref.~\cite{Fu:2017lps} that the quantum focusing conjecture is violated. However, in Ref.~\cite{Leichenauer:2017bmc}, it is pointed out that the analysis in Ref.~\cite{Fu:2017lps} can not be valid once the quantum expansion is smeared over a length scale of the order of the scale determined by the coupling constant of the Gauss-Bonnet gravity. The purpose of this paper is to evaluate the quantum expansion smeared over a length scale in the general quadratic gravity, which includes the Gauss-Bonnet gravity as a special choice of the parameters, and derive the condition for the averaging scale so that the quantum focusing conjecture holds for arbitrary spacetime.

This paper is organized as follows.
In Sec.~\ref{QFC}, we review the quantum focusing conjecture. Then, in Sec.~\ref{not smeared QFC}, we evaluate the quantum expansion and its derivative for the quadratic gravity. There we will find that the expressions reduce to those of the Gauss-Bonnet gravity studied in Refs.~\cite{Fu:2017lps, Leichenauer:2017bmc}. In Sec.~\ref{smeared QFC}, we evaluate the derivative of the smeared quantum expansion and derive the condition for the averaging scale so that the quantum focusing conjecture holds for arbitrary spacetime. In Sec.~\ref{example}, we investigate an axially symmetric surface in the five dimensional Schwarzschild spacetime as an example of the detailed calculation of the smeared quantum expansion. There we see that the quantum focusing conjecture is satisfied after the averaging procedure.
The final section~\ref{summary} is devoted to the summary and discussion.
We present the detailed calculations used in the main sections in the appendixes.  

Throughout the paper we mainly consider spacetime with the dimension $d \geq 5$, though we will comment for $d = 4$ case at some points.
We use indices $\mu,\nu,\rho,\sigma,\dots$ to denote the components of the tensor with respect to the (unspecified) coordinates $\{x^{\mu}\}$ in $d$-dimensional spacetime, while we use $i,j,k, \dots$ for the coordinates $\{y^{i}\}$ in co-dimension 2 surface. We use the unit $c = 1$ and $k_{\mathrm{B}} = 1$, where $c$ is the speed of light and $k_{\mathrm{B}}$ is the Boltzmann constant respectively.

\section{Review on the Quantum Focusing Conjecture}
\label{QFC}
In this section, we review the quantum focusing conjecture proposed by Ref.~\cite{Bousso:2015mna}.
The quantum focusing conjecture claims that the converging property of the null geodesics in classical General Relativity is promoted to a property of the generalized entropy in a semi-classical level just by replacing the expansion of the null geodesics $\theta$ with the quantum expansion $\Theta$ defined below.  The setup which we focus on is as follows.
Let $({\cal M}, g)$ be a (part of) spacetime that we are interested in. Let $\sigma$ be a compact, spacelike co-dimension 2 surface. Let $k^{\mu}$ be a null vector field on $\sigma$ which is orthogonal to $\sigma$.
By solving geodesic equations towards $k^{\mu}$ direction with the initial condition where the tangent vector reduces $k^{\mu}$ on $\sigma$, we can define the null hyper surface $N$ generated by these null geodesics. We define vector field $k^{\mu}$ on $N$ by the tangent vector of the affine parametrized null generator of $N$.
Let $ y = \{y^{i}\}$ be coordinates on $\sigma$ and $\lambda$ be the affine parameter of null generator of $N$. Then $(\lambda,\, y^{i})$ can be regarded as coordinates on $N$. We can identify a function $V$ of $y$ with the slice of $N$ defined by $\lambda = V(y)$. We call both the function and the surface by the same name. For example, we assume the surface $\sigma$ is characterized by a function named $\sigma$: $\lambda = \sigma(y)$.
Now we can describe the variation of a slice of $N$ by a variation of a function $V$.
The key insight is that the expansion of the null generator, $\theta := \nabla_{\mu} k^{\mu}$, at a point $(\lambda = V(y), y)$ can be expressed as the functional derivative of the area functional $A[V] := \int_{V} d^{d-2} y \sqrt{h(y)}$,
\begin{align}
 \theta(y) = \frac{1}{\sqrt{h(y)}}\frac{\delta A[V]}{\delta V(y)}, \label{cexpansion}
\end{align}
where $h$ is the determinant of the induced metric on the surface $V$.
See the appendix \ref{sec:functionalderivative}, in particular Eq.~\eqref{dAdV}, for the detailed calculation of the functional derivative.
We regard $\theta$ as a functional of $V$ as well as a function of coordinates $y$, such as $\theta = \theta[V;y]$, though we simply express it as $\theta(y)$ as long as there is no confusion.  In addition, we can express the converging property of the null geodesic congruence by the functional derivative of $\theta$, 
\begin{align}
 \frac{\delta \theta(y)}{\delta V(y_{1})} = \frac{d \theta}{d \lambda} \delta^{d-2} (y - y_{1}) \leq 0 ,\label{cfocusing}
\end{align}
which is ensured by the null energy condition because of the Raychaudhuri equation,
\begin{align}
 \frac{d \theta}{d \lambda} = - \frac{1}{d-2} \theta^2 - \sigma_{\mu\nu} \sigma^{\mu\nu} - R_{\mu\nu}k^{\mu} k^{\nu},\label{Raychaudhuri}
\end{align}
with $\sigma_{\mu\nu}$ being the shear tensor.
The right hand side is negative or zero when the null energy condition is satisfied,  $R_{\mu\nu} k^{\mu} k^{\nu} \geq 0$.

The quantum expansion~\cite{Bousso:2015mna} is defined by replacing the area functional $A$ in Eq.~\eqref{cexpansion} with the generalized entropy as
\begin{align}
 A[V] \rightarrow 4 G \hbar S_{\rm gen}[V].
\end{align}
Thus, the quantum expansion $\Theta$ is defined by
\begin{align}
 \Theta[V;y] := \frac{4 G \hbar}{\sqrt{h(y)}}\frac{\delta S_{\text{gen}}[V]}{\delta V(y)}.
\end{align}
Here the generalized entropy $S_{\rm gen}[V]$ is the total entropy of the system outside
\footnote{
To define ``outside'' and ``inside'' appropriately, we need to assume that $({\cal M}, g)$ is a globally hyperbolic spacetime with a Cauchy surface $\Sigma$ and $V$ is a surface on $\Sigma$ which divides $\Sigma$ into two parts.
}
the surface $V$, which includes the entanglement entropy of the quantum fields $S_{\text{out}}[V]$ obtained by tracing out the inside of $V$, as well as the gravitational entropy $S_{\text{g}}[V]$, that is the geometrical counter terms required by the renormalizaion procedures. The quantum expansion reduces to the usual expansion of the null geodesics if $S_{\text{gen}}$ is the Bekenstein--Hawking entropy $A[V]/4 G \hbar$. Though the quantum expansion is proposed by motivating to characterize the property of the semi-classical physics, we would like to emphasize that the quantum expansion differs from the usual expansion even in the classical limit $\hbar \rightarrow 0$ because the gravitational entropy $S_{\text{g}}[V]$ depends on the higher curvature corrections in the classical gravitational action,
\begin{align}
 S_{\rm g}[V] = \frac{1}{4 G \hbar} \Bigl(  A[V] +  \text{higher derivative corrections}  \Bigr) + {\cal O}(\hbar^{0}),
\end{align} 
as we will discuss in detail in the next section.
In this paper, we focus on the contributions relevant in the classical limit $\hbar \rightarrow 0$. In other words, we focus only on the contribution from the gravitational entropy, that is, $S_{\text{gen}}[V] = S_{\text{g}}[V]$.

The quantum focusing conjecture \cite{Bousso:2015mna} states that the non-positivity of the functional derivative of the quantum expansion,
\begin{align}
 \frac{\delta \Theta[V;y]}{\delta V(y_{1})} \leq 0,\label{qfc}
\end{align}
which is an analogy of the converging property of null geodesics \eqref{cfocusing}.
The purpose of this paper is to clarify the relation between the validity of the quantum focusing conjecture and the cutoff scale of the gravitational theory with the higher curvature corrections.

\section{Gravitational Quantum Expansion with the Higher Derivative Corrections}
\label{not smeared QFC}
\subsection{Action and the Equations of Motion}
We consider the higher derivative corrections to the $d$-dimensional Einstein--Hilbert action at the fourth order in the derivative expansion, that is, the quadratic gravity.
The action is given by 
\begin{align}
\label{quadratic gravity}
I=\frac{1}{16\pi G}\int d^dx\sqrt{-g}[R+\alpha R^2+\beta R_{\mu\nu}R^{\mu\nu}+\gamma R_{\mu\nu\rho\sigma}R^{\mu\nu\rho\sigma}],
\end{align}
where $\alpha,\beta,\gamma$ are the renormalized coupling constants with the dimension of a length square. Since we focus on the classical limit $\hbar \rightarrow 0$, the running of these constants, as well as that of the Newton constant $G$, is ignored.

The equations of motion can be written as
\begin{align}
\label{motion of quadratic}
 \frac{16 \pi G}{\sqrt{-g}}\frac{\delta I}{\delta g^{\mu\nu}} 
&=
R_{\mu\nu} - \frac{1}{2} R g_{\mu\nu} + 
\alpha H_{\mu\nu}^{(1)} +\beta H_{\mu\nu}^{(2)} + \gamma H_{\mu\nu}^{(3)} = 0,
\end{align}
with 
\begin{align}
 H_{\mu\nu}^{(1)} &= 2 (R_{\mu\nu} - \nabla_{\mu} \nabla_{\nu} ) R - g_{\mu\nu} \left(\frac{1}{2} R^2 - 2 \Box R\right), \\
 H_{\mu\nu}^{(2)} &=  2 R_{\mu\rho\nu\sigma}R^{\rho\sigma} + \Box R_{\mu\nu} - \nabla_{\mu} \nabla_{\nu} R
- \frac{1}{2}g_{\mu\nu} \left( R_{\rho\sigma} R^{\rho\sigma} - \Box R \right) \\
 H_{\mu\nu}^{(3)} &= 2 R_{\mu\rho\sigma\tau}R_{\nu}{}^{\rho\sigma\tau} - \frac{1}{2} g_{\mu\nu} R_{\rho\sigma\tau\lambda} R^{\rho\sigma\tau\lambda} + 4R_{\mu\rho\nu\sigma}R^{\rho\sigma}-4R_{\mu\rho}R^{\rho}_{\ \nu}-2\nabla_{\mu}\nabla_{\nu}R+4\Box R_{\mu\nu} .
\end{align}
We solve these equations of motion perturbatively with respect to the parameters $\alpha,\beta,\gamma$. We express the order of the perturbation by a parameter $\epsilon$, for example,  $ \alpha = \mathcal{O}(\epsilon)$, $\gamma^2 = \mathcal{O}(\epsilon^2)$, and so on.
For the validity of perturbation, we assume $\sqrt{|\gamma|} \ll L$ denoting $L$ as the length scale of the Weyl tensor.
The Ricci tensors in the expression of $H_{\mu\nu}^{(1)}, H_{\mu\nu}^{(2)}, H_{\mu\nu}^{(3)}$ can be pushed to higher order in $\epsilon$. As a results, the contribution of the $\alpha$ and $\beta$ terms in the equations of motion disappear and the equations of motion can be simplified as 
\begin{align}
 R_{\mu\nu} = - 2 \gamma \left( C_{\mu\rho\sigma\tau}C_{\nu}{}^{\rho\sigma\tau}  - \frac{1}{2 (d-2)} g_{\mu\nu} C_{\rho\sigma\tau\lambda} C^{\rho\sigma\tau\lambda} \right)  + {\cal O}\left( \epsilon^2 \right). \label{EOM} 
\end{align}

Since only the parameter $\gamma$ appears in Eq.~\eqref{EOM}, 
the solution of the equation of motion at the leading order is independent of the choice of $\alpha$ and $\beta$. We can obtain the same solution if we start from the specific choice of the value of $\alpha$ and $\beta$, for example, the Gauss--Bonnet coupling, $\alpha = \gamma, \beta = - 4 \gamma$. We would like to emphasize that, however,
it is not obvious that the entropy, the quantum expansion and its functional derivative are independent of the choice of $\alpha, \beta$. For this reason, we will maintain the parameters $\alpha$ and $\beta$ general below. Note that the right hand sides of the equations of motion \eqref{EOM} vanish for $d = 4$ because of the identity $
C^{\mu\rho}{}_{\sigma\tau} C_{\nu\rho}{}^{\sigma\tau} - \frac{1}{4} \delta^{\mu}{}_{\nu} C^{\lambda\rho}{}_{\sigma\tau} C_{\lambda\rho}{}^{\sigma\tau} = - \frac{15 }{2} \delta^{[\mu}{}_{\nu} C^{\lambda\rho}{}_{\sigma\tau} C_{\lambda\rho}{}^{\sigma\tau]} = 0$ which holds in four dimension. 
This is because the Gauss--Bonnet term becomes total derivative and does not affect the equations of motion in $d = 4$. 

\subsection{Gravitational Entropy}
The higher derivative corrections of the gravitational action relates with that of the gravitational entropy. Here we will use the gravitational entropy proposed by Dong \cite{Dong:2013qoa} and Camps \cite{Camps:2013zua}, which motivated to describe the gravitational dual of the entanglement entropy of CFT in the AdS/CFT setup (see also \cite{Fursaev:2013fta,Bhattacharyya:2013jma,Alishahiha:2013zta, Bhattacharyya:2013gra,Bhattacharyya:2014yga,Miao:2014nxa,Dong:2015zba}).
Let us consider a co-dimension 2 surface $\sigma$ and the null hyper surface $N$ generated by the null generators with the tangent vector $k^{\mu}$, and the coordinates $(\lambda, y)$ on $N$ as is the previous section.
For a given slice $V$ of the null hyper surface $N$, we can introduce another null vector field $l^{\mu}$ on $V$ that is orthogonal to $V$ and satisfies
\begin{align}
 g_{\mu\nu} l^{\mu} l^{\nu} = 0, \qquad g_{\mu\nu} k^{\mu} l^{\nu} = -1.
\end{align} 
Then the induced metric $h_{\mu\nu}$ on $V$ can be defined as 
\begin{align}
h_{\mu\nu} :=  g_{\mu\nu} + k_{\mu} l_{\nu} + l_{\mu} k_{\nu}.
\end{align}
$h_{\mu}{}^{\nu}$ becomes the projection tensor to the co-dimension 2 surface $V$.  Similarly, $q_{\mu}{}^{\nu} := - k_{\mu} l^{\nu} - l_{\mu} k^{\nu}$ defines the projection tensor to the 2-dimensional space orthogonal to $V$.
We define the second fundamental tensor $K_{\rho\mu\nu}$ by
\begin{align}
 K_{\rho\mu\nu} := h_{\mu}{}^{\sigma} h_{\nu}{}^{\tau} \nabla_{\sigma} q_{\tau\rho} = - \left( h_{\mu}{}^{\sigma} h_{\nu}{}^{\tau} \nabla_{\sigma} k_{\tau} \right) l_{\rho} - \left( h_{\mu}{}^{\sigma} h_{\nu}{}^{\tau} \nabla_{\sigma} l_{\tau} \right) k_{\rho},
\end{align}
and write
\begin{align}
 K^{(X)}_{\mu\nu} := X^{\rho} K_{\rho\mu\nu}
\end{align}
for any vector $X^{\mu}$. For example, 
\begin{align}
 K^{(k)}_{\mu\nu} = h_{\mu}{}^{\sigma} h_{\nu}{}^{\tau} \nabla_{\sigma} k_{\tau}, \qquad  K^{(l)}_{\mu\nu} = h_{\mu}{}^{\sigma} h_{\nu}{}^{\tau} \nabla_{\sigma} l_{\tau},\label{defKkKl}  
\end{align} 
and so on. See, e.g., Ref.~\cite{Cao:2010vj} for more details though our notation is slightly different. Note that $K_{\rho\mu\nu} = q_{\rho}{}^{\sigma} K_{\rho\mu\nu} = h_{\mu}{}^{\sigma} K_{\rho\sigma\nu} = h_{\nu}{}^{\sigma} K_{\rho\mu\sigma}$. $K^{\rho} :=  K^{\rho}{}_{\mu}{}^{\mu}$ is called mean curvature vector. We also use the notation $K^{(X)} := X^{\rho} K_{\rho}$.
Then, the Dong -- Camps gravitational entropy of the quadratic gravity is given by
\begin{align}
 \label{entropy of quadratic gravity}
 S_{\text{g}}[V] = \frac{A[V]}{4 G \hbar} + 
\frac{1}{4 G \hbar} \int_{V} d^{d-2} y \sqrt{h} & \biggl[ 2 \alpha R + \beta \left(q^{\mu\nu} R_{\mu\nu} - \frac{1}{2} K_{\rho} K^{\rho}\right) \notag\\
&\qquad  + 2 \gamma \left(q^{\mu\nu}q^{\rho\sigma}R_{\mu \rho}{}_{\nu \sigma} - K_{\mu\nu\rho} K^{\mu\nu\rho} \right) \biggr].
\end{align}
Since we assume $R_{\mu\nu} = \mathcal{O}(\epsilon)$, the gravitational entropy can be expressed as
\begin{align}
 S_{\text{g}}[V] &= \frac{A[V]}{4 G \hbar} + 
\frac{1}{4 G \hbar} \int_{V} d^{d-2} y \sqrt{h} \left[  - \frac{1}{2} \beta K_{\rho} K^{\rho} + 2 \gamma \left(q^{\mu\nu}q^{\rho\sigma}R_{\mu \rho}{}_{\nu \sigma} - K_{\mu\nu\rho} K^{\mu\nu\rho} \right) + {\cal O}(\epsilon^2)\right].
\end{align}
The contribution from the $\alpha$ term is dropped off, while that from the $\beta$ term still remains.
In addition, by eliminating $q^{\mu\nu}q^{\rho\sigma} R_{\mu\rho\nu\sigma}$ by the Gauss equation \eqref{GausseqRS} in the appendix \ref{sec:A}, we can express the gravitational entropy as an integral of the covariant quantity on $\sigma$:
\begin{align}
 S_{\text{g}}[V] & = \frac{A[V]}{4 G \hbar} + 
\frac{1}{4 G \hbar} \int_{V} d^{d-2} y \sqrt{h} \left[  (\beta + 4 \gamma) K^{(l)} K^{(k)} + 2 \gamma {}^{(d - 2)} R + {\cal O}(\epsilon^2)\right],  \label{entropy of quadratic gravity_2}
\end{align}
where ${}^{(d-2)} R$ is the $d-2$ dimensional Ricci scalar associated with $h_{\mu\nu}$ on $V$. 

\subsection{Quantum Expansion and Its Derivative}
\label{sec:QE}
The purpose of this paper is to see the validity of quadratic gravity from the view point of the quantum focusing conjecture. For this purpose, let us consider a specific point $p$ on the surface $\sigma$ where the leading order contribution of the quantum expansion, that is the usual expansion $\theta$, saturates the equality of the quantum focusing conjecture \eqref{qfc}. Since it reduces to the expression by the classical expansion \eqref{cfocusing}, the equality holds if $\theta|_{p} = 0$, $\sigma_{\mu\nu}|_{p} = 0$.
 Recall that the expansion $\theta$ and shear tensor $\sigma_{\mu\nu}$ of the null generator of $N$ on the surface $\sigma$ can be expressed as 
\begin{align}
  \theta = K^{(k)}, \qquad \sigma_{\mu\nu} = K^{(k)}_{\mu\nu} - \frac{1}{d-2} K^{(k)} h_{\mu\nu}. \label{defthetasigma}
\end{align}
Then our requirement is the presence of a point $p$ on $\sigma$, which satisfies
\begin{align}
K_{\mu\nu}^{(k)}\mid_p = \mathcal{O}(\epsilon). \label{extrinsic curvature}
\end{align}

Let us first confirm that any term $\Delta S[V]$ of the following form in the entropy formula does not contribute to the quantum expansion and its derivative at the point $p$:
\begin{align}
 \Delta S[V] := \frac{\gamma}{4 G \hbar} \int_{V} d^{d-2} y \sqrt{h} F K^{(k)}, \label{DeltaS}
\end{align}
where $F$ is arbitrary scalar function constructed from the covariant quantities on $V$.
For example, in our entropy formula~\eqref{entropy of quadratic gravity_2}, $F$ can be understood as $F = (4 + \beta/ \gamma) K^{(l)}$.
The contribution from $\Delta S[V]$ to the quantum expansion, say $\Delta \Theta$, can be evaluated by using Eq.~\eqref{LieK} in the appendix as  
\begin{align}
\Delta \Theta[V; y] &:=  \frac{4 G \hbar}{\sqrt{h}} \frac{\delta \Delta S[V]}{\delta V(y)} \notag\\
&= \gamma \left( \frac{1}{\sqrt{h}} \mathsterling_{k}(\sqrt{h} F)  K^{(k)}  + F \mathsterling_{k} K^{(k)} \right)  \notag\\
& =
\gamma \left( \frac{1}{\sqrt{h}} \mathsterling_{k}(\sqrt{h} F)  K^{(k)} - F K^{(k)}{}^{\mu\nu} K^{(k)}_{\mu\nu} \right) + \mathcal{O}\left( \epsilon^2 \right)
\label{dKdV}
\end{align}
and hence, denoting the $y$-coordinate value of the point $p$ as $y_{p}$, we obtain,
\begin{align}
\Delta \Theta[\sigma; y_{p}] = \mathcal{O} \left( \epsilon^2 \right).
\end{align}
This means that any term proportional to $K^{(k)}$ in the entropy formula does not contribute to the value of the quantum expansion at the point $p$.
Let us, then, investigate the first derivative of $\Delta \Theta$ at the point $p$. Since the first term in Eq.~\eqref{dKdV} is again proportional to $K^{(k)}$, by repeating above discussion, one can see that the functional derivative of this term vanishes at $p$. In addition, since the second term in Eq.~\eqref{dKdV} is quadratic in $K^{(k)}_{\mu\nu}$, the functional derivative of this term also vanish at the point $p$.
Thus, functional derivative of $\Delta \Theta$ can be evaluated as 
\begin{align}
\left.  \frac{\delta \Delta \Theta[V;y]}{\delta V(y_{1})} \right|_{V = \sigma, y = y_{p}} = \mathcal{O}(\epsilon^2).
\end{align}
Thus, we find that the functional derivative of the quantum expansion, as well as the quantum expansion itself, does not depend on the terms $\Delta S$ in the entropy formula. In particular, there is no contribution of $\beta$ term, as well as $\alpha$ term, to the functional derivative of the quantum expansion at the point $p$.
As a result, the expressions of the quantum expansion and its derivative at the point $p$, which we will evaluate below, are equivalent for any choice of $\alpha$ and $\beta$. Especially, our expression reduces to that for the Gauss-Bonnet coupling $\alpha = \gamma$ and $\beta = - 4 \gamma$, which is investigated in Refs.~\cite{Fu:2017lps} and \cite{Leichenauer:2017bmc}.
We would like to emphasize that, though we start from the Dong--Camps entropy formula \eqref{entropy of quadratic gravity}, our calculations below are applicable for any gravitational entropy which differ by the terms proportional to $K^{(k)}$ when $R_{\mu\nu} \sim \mathcal{O}(\epsilon)$, such as Wald's entropy formula. 

To summarize, the relevant terms in the entropy formula for the calculation of the derivative of the quantum expansion at the point $p$ can be expressed as 
\begin{align}
 S_{\text{g}}[V] & = \frac{A[V]}{4 G \hbar} + 
\frac{ 2 \gamma}{4 G \hbar} I^{\text{EH}}[V] + \Delta S[V], \label{Gauss-Bonnet entropy}
\end{align}
as is the Gauss--Bonnet case.
Here $I^{\text{EH}}[V]$ is $d-2$ dimensional Einstein--Hilbert action defined by
\begin{align}
I^{\text{EH}}[V] := \int_{V} d^{d-2} y \sqrt{h} \, {}^{(d - 2)} R. 
\end{align}
Following the analysis for the Gauss--Bonnet case in Refs.~\cite{Fu:2017lps, Leichenauer:2017bmc}, the quantum expansion associated with the quadratic curvature terms can be evaluated as follows: 
\begin{align}
\Theta[V;y]&=\frac{4 G \hbar}{\sqrt{h}}\frac{\delta S_{\text{g}}[V]}{\delta V(y)}\notag\\
&=\frac{1}{\sqrt{h}} \frac{\delta A[V]}{\delta V(y)} + 2 \gamma \frac{1}{\sqrt{h}} \frac{\delta I^{\text{EH}}[V]}{\delta V(y)} + \gamma\, \mathcal{O}(K^{(k)}, K^{(k)}_{\mu\nu} K^{(k)}{}^{\mu\nu}) + \mathcal{O}\left( \epsilon^2 \right) \notag\\  
&= \theta - 4\gamma\ {}^{(d-2)}G^{\mu\nu}K^{(k)}_{\mu\nu} + \gamma\, \mathcal{O}(K^{(k)}, K^{(k)}_{\mu\nu} K^{(k)}{}^{\mu\nu}) + \mathcal{O}\left( \epsilon^2 \right),
\end{align}
where $^{(d-2)}G_{\mu\nu}$ is $(d-2)$ dimensional Einstein tensor and the functional derivative of the Einstein--Hilbert action is given by Eq.~\eqref{dGdV} in the appendix.

Since the quantum expansion is expressed in terms of the covariant quantity on $V$, the functional derivative of the quantum expansion can be calculated by Lie derivative, 
\begin{align}
\frac{\delta \Theta[V;y]}{\delta V(y_1)} &=
\frac{d \Theta}{d \lambda} \, \delta^{(d-2)}(y - y_{1}),
\end{align}
where
\begin{align}
 \frac{d \Theta}{d \lambda} = \mathsterling_{k} \Theta = \frac{d \theta}{d \lambda} - 4 \gamma\, {}^{(d-2)} G^{\mu\nu} \mathsterling_{k} K^{(k)}_{\mu\nu} + \gamma\,\mathcal{O}\left(K^{(k)}_{\mu\nu}\right) + \mathcal{O}\left( \epsilon^2 \right)
.
\end{align}
The statement of the quantum focusing conjecture can be simply expressed as $d \Theta/d \lambda \leq 0$.
By using the Raychaudhuri equation \eqref{Raychaudhuri} and the equation of motion (\ref{EOM}), the contribution from the classical expansion $\theta$ can be calculated as
\begin{align}
 \frac{d \theta}{d \lambda} &= - \frac{1}{d-2} \theta^2 - \sigma_{\mu\nu} \sigma^{\mu\nu} - R_{\mu\nu} k^{\mu} k^{\nu} \notag\\ 
&= - \frac{1}{d-2} \theta^2 - \sigma_{\mu\nu} \sigma^{\mu\nu} +  2 \gamma \, C_{\mu\rho\sigma\tau} C_{\nu}{}^{\rho\sigma\tau} k^{\mu} k^{\nu} + {\cal O}\left( \epsilon^2 \right) \notag\\
&= - \frac{1}{d-2} \theta^2 - \sigma_{\mu\nu} \sigma^{\mu\nu} +  2 \gamma \,
\left( 
\mathcal{C}_{\boldsymbol{k}\mu\nu\rho} \mathcal{C}_{\boldsymbol{k}}{}^{\mu\nu\rho} - 2 \mathcal{C}_{\boldsymbol{k}}{}^{\mu}{}_{\nu\mu} \mathcal{C}_{\boldsymbol{k}}{}^{\rho \nu}{}_{\rho}
-4 \mathcal{C}_{\boldsymbol{k} \mu \boldsymbol{k} \nu} \mathcal{C}_{\boldsymbol{k}}{}^{\mu}{}_{\boldsymbol{l}}{}^{\nu} \right)  + {\cal O}\left( \epsilon^2 \right),\label{classicalthetadot}
\end{align}
where we use the identity \eqref{CkCk} in the appendix. Here we used the notion
\begin{align}
 \mathcal{C}_{\boldsymbol{k}\mu\nu\rho} &:= k^{\alpha} h_{\mu}{}^{\beta} h_{\nu}{}^{\gamma} h_{\rho}{}^{\delta} C_{\alpha\beta\gamma\delta}, \\
 \mathcal{C}_{\boldsymbol{k} \mu \boldsymbol{k} \rho} &:= k^{\alpha} h_{\mu}{}^{\beta} k^{\gamma} h_{\rho}{}^{\delta} C_{\alpha \beta \gamma \delta}, \\
 \mathcal{C}_{\boldsymbol{k} \mu \boldsymbol{l} \rho} &:= k^{\alpha} h_{\mu}{}^{\beta} l^{\gamma} h_{\rho}{}^{\delta} C_{\alpha \beta \gamma \delta}.
\end{align}
On the one hand, the contribution from the higher curvature corrections in the entropy formula can be evaluated as
\begin{align}
 - 4 \gamma {}^{(d-2)}G^{\mu\nu} \mathsterling_{k} K^{(k)}_{\mu\nu} &= 4 \gamma\, {}^{(d-2)}G^{\mu\nu} R_{\rho\mu\sigma\nu}k^{\rho}k^{\sigma}+ \gamma\, \mathcal{O} \left(K^{(k)}_{\mu\nu}\right)  \notag\\
&= 2 \gamma * 4 \mathcal{C}_{\boldsymbol{k} \mu \boldsymbol{k} \nu} \mathcal{C}_{\boldsymbol{k}}{}^{\mu}{}_{\boldsymbol{l}}{}^{\nu} + \gamma \, \mathcal{O} \left(K^{(k)}_{\mu\nu}\right) + {\cal O}\left( \epsilon^2 \right)
 ,\label{LieGK}
\end{align}
where we used Eq.~\eqref{LieKmunu} in the first equality and Eq.~\eqref{GRkk} in the second equality.
Combining Eqs.~\eqref{classicalthetadot} and \eqref{LieGK}, we obtain
\begin{align}
\frac{d \Theta}{d \lambda}
&=
- \frac{1}{d-2} \theta^2 - \sigma_{\mu\nu} \sigma^{\mu\nu}
+ 2\gamma\left(
\mathcal{C}_{\boldsymbol{k}\mu\nu\rho} \mathcal{C}_{\boldsymbol{k}}{}^{\mu\nu\rho} - 2 \mathcal{C}_{\boldsymbol{k}}{}^{\mu}{}_{\nu\mu} \mathcal{C}_{\boldsymbol{k}}{}^{\rho \nu}{}_{\rho}
 \right) 
+ \gamma\, \mathcal{O} \left( K^{(k)}_{\mu\nu} \right)
+ \mathcal{O}\left(\epsilon^2\right).\label{QexpressedC}
\end{align}

In the final expression~\eqref{QexpressedC}, the dependence of the spacetime curvature appears only through the specific components of the Weyl tensor, $\mathcal{C}_{\boldsymbol{k}\mu\nu\rho}$.
Following Refs.~\cite{Fu:2017lps, Leichenauer:2017bmc} (see also Ref.~\cite{Coley:2009is}), we express the independent components of $\mathcal{C}_{\boldsymbol{k}\mu\nu\rho}$ as
\begin{align}
 v_{\nu} := - \frac{1}{d - 3} \mathcal{C}_{\boldsymbol{k}\mu \nu}{}^{\mu},\qquad 
 T_{\mu\nu\rho} := 
\mathcal{C}_{\boldsymbol{k}\mu\nu\rho} + \frac{2}{d-3} h_{\mu[\nu|} \mathcal{C}_{\boldsymbol{k} \sigma |\rho]}{}^{\sigma}
,
\end{align}
or
\begin{align}
 \mathcal{C}_{\boldsymbol{k} \mu\nu\rho} = 2 h_{\mu[\nu}v_{\rho]} + T_{\mu\nu\rho},\label{Ck=v+T}
\end{align}
where $v_{\mu}$ and $T_{\mu\nu\rho}$ are tensor on the co-dimension 2 space, that is, they satisfy $v_{\mu} = h_{\mu}{}^{\nu} v_{\nu}$ and $T_{\mu\nu\rho} = h_{\mu}{}^{\sigma} T_{\sigma\nu\rho} = h_{\nu}{}^{\sigma} T_{\mu\sigma\rho} = h_{\rho}{}^{\sigma} T_{\mu\nu\sigma}$. In addition, $T_{\mu\nu\rho}$ has the following symmetry on the indices,  
\begin{align}
\label{T}
 T_{[\mu\nu\rho]} = T_{\mu(\nu\rho)} = T^{\mu}{}_{\nu\mu} = 0.
\end{align}
$(d - 3)(d-2)(d-1)/3$ independent components of $\mathcal{C}_{\boldsymbol{k} \mu\nu\rho}$ are described by $d-2$ independent components of $v_{\mu}$ and $d(d-2)(d-4)/3$ independent components of $T_{\mu\nu\rho}$. Note that $T_{\mu\nu\rho}$ identically vanish for $d=4$. By using $v_{\mu}$ and $T_{\mu\nu\rho}$, the Weyl tensor terms in Eq.~\eqref{QexpressedC} can be expressed as
\begin{align}
 \mathcal{C}_{\boldsymbol{k} \mu\nu\rho} \mathcal{C}_{\boldsymbol{k}}{}^{\mu\nu\rho} = 2 (d-3) v_{\mu} v^{\mu} + T_{\mu\nu\rho} T^{\mu\nu\rho}, \qquad \mathcal{C}_{\boldsymbol{k}}{}^{\mu}{}_{\nu \mu} \mathcal{C}_{\boldsymbol{k}}{}^{\rho}{}^{\nu} {}_{\rho} = (d-3)^2 v_{\mu} v^{\mu},
\end{align}
and hence we obtain
\begin{align}
 \frac{d \Theta}{d \lambda} &=
- \frac{1}{d-2} \theta^2 - \sigma_{\mu\nu} \sigma^{\mu\nu}
+
2 \gamma \Bigl( - 2 (d-3)(d-4) v_{\mu} v^{\mu} + T_{\mu\nu\rho} T^{\mu\nu\rho} \Bigr) + \gamma\, \mathcal{O} \left( K^{(k)}_{\mu\nu} \right) + \mathcal{O}\left( \epsilon^2 \right)
. \label{QexpressedvT}
\end{align}
Note that the contributions from the $\gamma$ term disappear for $d = 4$. 

\section{Validity of the Quantum Focusing Conjecture and the Cutoff Scale}
\label{smeared QFC}
\subsection{Quantum Expansion at the Point $p$}
Let us see the sign of $d \Theta/ d \lambda$ at the point $p$ defined by Eq~\eqref{extrinsic curvature}. From the expression \eqref{QexpressedvT}, we obtain
\begin{align}
\label{the derivative QE}
\left. \frac{d \Theta}{d \lambda} \right|_p 
= 
2 \gamma  \left. \Bigl( - 2 (d-3)(d-4) v_{\mu} v^{\mu} + T_{\mu\nu\rho} T^{\mu\nu\rho} \Bigr) \right|_{p} + \mathcal{O}\left(\epsilon^2 \right).
\end{align}
Note that the expansion and the shear are $\mathcal{O}(\epsilon)$ at $p$ and appear as a square in the derivative of the quantum expansion. Therefore, they do not contribute in the analysis of $\mathcal{O}(\epsilon)$. The quantum focusing conjecture states that Eq. (\ref{the derivative QE}) is nonpositive. However, as discussed in Ref.~\cite{Fu:2017lps} for the Gauss-Bonnet case, the sign of Eq. (\ref{the derivative QE}) depends on the value of the independent components of the Weyl tensor, $v_{\mu}$ and $T_{\mu\nu\rho}$. Thus, there is a point $p$ where the quantum focusing conjecture is not satisfied in the general situation. For example, in the case of $\gamma > 0$, Eq. (\ref{the derivative QE}) is positive for $v_{\mu} = 0$ and $T_{\mu\nu\rho} \neq 0$ and in the case of $\gamma < 0$, Eq. (\ref{the derivative QE}) is positive for $v_\mu \neq 0$ and $T_{\mu\nu\rho}=0$. 

\subsection{Smeared Quantum Expansion around the Point $p$}
The result in the previous section, as is the result in Ref.~\cite{Fu:2017lps} for Gauss-Bonnet case, reveals the potential violation of the quantum focusing conjecture at the order $\mathcal{O}( \epsilon )$, due to the higher derivative corrections to the gravitational action and the gravitational entropy. However, such a result will be altered once the cutoff scale of the theory is taken into account, as suggested in Ref.~\cite{Leichenauer:2017bmc}. 
In Ref.~\cite{Leichenauer:2017bmc}, it was shown that the discussion in Ref.~\cite{Fu:2017lps} can not be applied if the quantum expansion is smeared around the point $p$. Concretely, it was shown that, if we consider the smearing scale $\ell$ is of the order of $ \sqrt{|\gamma|}$, the contribution of the smearing does not vanish at $\mathcal{O}(\epsilon)$, when $d \Theta/ d\lambda|_{p}$ does not vanish at $\mathcal{O}(\epsilon)$. 
In this section, we 
evaluate the $\mathcal{O}(\epsilon)$ contribution by the smearing rigorously and derive the sufficient condition for the smearing scale $\ell$ to satisfy the quantum focusing conjecture.

To formulate the smearing rigorously, let us use the Riemannian normal coordinates~\cite{Berline1992HeatKA}, $\{y^{i} \} = \{y^1,y^2,y^3,\cdots,y^{d-2}\}$, on $\sigma$ around the point $p$ defined by Eq.~\eqref{extrinsic curvature}.
The metric on $\sigma$ can be expanded as
\begin{align}
h_{ij} dx^i dx^j=\delta_{ij} dy^i dy^j - \frac{1}{3}{}^{(d-2)}R_{ikjl}|_{p}\, y^k y^l dy^i dy^j+\cdots,
\end{align}
where $\delta_{ij}$ is Kronecker's delta. The point $p$ corresponds to $y^{i} = 0$.
Since ${}^{(d-2)}R_{ikjl}|_{p}$ is $\mathcal{O}(L^{-2})$ because of the Gauss equation \eqref{GaussEquation}, the Riemannian normal coordinates provide a local inertial frame for $y_{i} \leq \ell \sim \sqrt{|\gamma|} \ll L$.
We can approximate the derivative of the quantum expansion \eqref{QexpressedvT} in their Taylor series around a point $p$.  Then, we obtain
\begin{align}
\frac{d \Theta}{d \lambda} &= \left. \frac{d \Theta}{d \lambda} \right|_{p} - \frac{1}{d-2} y^{i} y^{j} \left. \left( \partial_{i} \theta \partial_{j} \theta \right) \right|_{p} - y^{i} y^{j}
\left. \left(\partial_{i} \sigma_{kl} \partial_{j} \sigma^{kl} \right) \right|_{p} + {\cal O}\left( \epsilon^{3/2} \right),
\end{align} 
where we regard $y^{i} < \ell$ and $\ell$ is $\mathcal{O}(\sqrt{\epsilon})$ in our perturbative expansion.
Next, we take the average of $d \Theta/ d \lambda$ over the $(d-2)$ dimensional hypercube in the Riemannian normal coordinates with a side length $\ell$. We define the average of a function $f(y)$ by
\begin{align}
 \braket{f}_{\ell} := \frac{1}{\ell^{d-2}} \int_{- \ell/2}^{\ell/2} d^{d-2} y f(y) .
\end{align}
By using the property
\begin{align}
\braket{1}_{\ell} = 1, \qquad  \braket{y^{i} y^{j}}_{\ell} = \frac{\ell^2}{12} \delta^{ij},
\end{align}
we obtain
\begin{align}
 \left\langle \frac{d \Theta}{d \lambda} \right\rangle_{\ell} &  = \left. \frac{d \Theta}{d \lambda} \right|_{p} - \frac{\ell^2}{12} \left. \left( \frac{1}{d-2}  \partial_{i} \theta \partial^{i} \theta + \partial_{i} \sigma_{jk} \partial^{i} \sigma^{jk}   \right) \right|_{p} + \mathcal{O}\left( \epsilon^2 \right).
\end{align}
Note that $\mathcal{O}(\epsilon^{3/2})$ terms disappear by the smearing because they are linear or cubic in $y^{i}$. 
By translating the result in the Riemannian coordinates to the general coordinates, we obtain
\begin{align}
 \left\langle \frac{d \Theta}{d \lambda} \right\rangle_{\ell} &  = \left. \frac{d \Theta}{d \lambda} \right|_{p} - \frac{\ell^2}{12} \left. \left( \frac{1}{d-2} D_{\mu} \theta D^{\mu} \theta + D_{\mu} \sigma_{\nu\rho} D^{\mu} \sigma^{\nu\rho}  \right)\right|_{p} +\left(\epsilon^2 \right) ,
\label{average Q}
\end{align}
where $D_{\mu}$ is the covariant derivative associated with $h_{\mu\nu}$.
We express the independent components of $D_{\mu} \theta$ and $D_{[\mu} \sigma_{\nu]\rho}$ by
\begin{align}
v^{(\theta)}_{\mu} &:= \frac{1}{d-2} D_\mu \theta, \label{vtheta} \\
v^{(\sigma)}_{\mu} &:=  - \frac{1}{d-3} D_{\nu}\sigma_{\mu}{}^{\nu}, \label{vsigma} \\
T^{(\sigma)}_{\rho\mu\nu} &:= -2 \left( D_{[\mu} \sigma_{\nu]\rho} + \frac{1}{d-3} D_{\alpha}\sigma_{[\mu}{}^{\alpha} h_{\nu]\rho} \right) \label{Tsigma}.
\end{align}
In other words, we express $D_{\mu} \theta$ and $D_{\mu} \sigma_{\nu\rho}$ as 
\begin{align}
 D_{\mu} \theta &= (d-2) v^{(\theta)}_{\mu},\qquad  D_{\mu} \sigma_{\nu\rho} = D_{(\mu} \sigma_{\nu)\rho} + v^{(\sigma)}_{[\mu} h_{\nu] \rho} - \frac{1}{2} T^{(\sigma)}_{\rho\mu\nu}.
\end{align}
In this notion, the Codazzi equations \eqref{Codazzi} can be simply expressed as 
\begin{align}
 v^{(\theta)}_{\mu} |_{p} + v^{(\sigma)}_{\mu}|_{p} = v_{\mu} |_{p}, \qquad 
 T_{\rho\mu\nu}^{(\sigma)} |_{p} = T_{\rho\mu\nu} |_{p}.
\end{align}

By rewriting the independent components of the Weyl tensor by $v^{(\theta)}_{\mu}, v^{(\sigma)}_{\mu}$ and $T^{(\sigma)}_{\mu\nu\rho}$,
the derivative of the smeared quantum expansion can be expressed as
\begin{align}
  \left\langle \frac{d \Theta}{d\lambda} \right\rangle_{\ell} &= - \frac{\ell^2}{12}  \biggl( (d-2) v_\mu^{(\theta)} v^{(\theta) \mu} + \frac{d-3}{2} v_\mu^{(\sigma)} v^{(\sigma) \mu} \notag\\
& \qquad \qquad \qquad + 48 \frac{\gamma}{\ell^2} (d-3)(d-4) (v_\mu^{(\theta)} + v_\mu^{(\sigma)})(v^{(\theta) \mu} + v^{(\sigma)\mu}) \left. \biggr) \right|_{p} \notag\\
& \quad -  \frac{\ell^2}{48}\left( 1- 96 \frac{\gamma}{\ell^2} \right) \left. \left( T_{\rho\mu\nu}^{(\sigma)} T^{(\sigma) \rho\mu\nu}\right) \right|_{p}- \frac{\ell^2}{12}  \left.\left(D_{(\mu} \sigma_{\nu)\rho} D^{(\mu} \sigma^{\nu)\rho}\right)\right|_{p} + \mathcal{O}(\epsilon^2) .
\end{align}
By introducing the $2 \times 2$-matrix
\begin{align}
 M_{IJ} :=
\begin{pmatrix}
 d-2 + 48 (d - 3)(d - 4) \frac{\gamma}{\ell^2} & 48 (d - 3)(d - 4) \frac{\gamma}{\ell^2}\\
  48 (d - 3)(d - 4) \frac{\gamma}{\ell^2} & \frac{d-3}{2} + 48 (d - 3)(d - 4) \frac{\gamma}{\ell^2}
\end{pmatrix},
\end{align}
and $2$-component vector
\begin{align}
 v_{\mu}^{I} := 
\begin{pmatrix}
 v_{\mu}^{(\theta)}\\
 v_{\mu}^{(\sigma)}
\end{pmatrix}, 
\end{align}
the final result can be expressed as
\begin{align}
\label{smeared Q}
 \left\langle \frac{d \Theta}{d \lambda} \right\rangle_{\ell}
&= - \frac{\ell^2}{12} M_{IJ} \left. \left(v^{I}_{\mu} v^{J}{}^{\mu}\right)\right|_{p} - \frac{\ell^2}{48} \left(1 - 96 \frac{\gamma}{\ell^2}\right) \left. \left(T^{(\sigma)}_{\rho\mu\nu} T^{(\sigma)\rho\mu\nu}\right) \right|_{p} \notag\\
&\quad - \frac{\ell^2}{12} \left. \left(D_{(\mu}\sigma_{\nu)\rho} D^{(\mu}\sigma^{\nu)\rho} \right) \right|_{p} + \mathcal{O}(\epsilon^2).
\end{align}
Thus, the quantum focusing conjecture is satisfied for any choice of $v^{(\theta)}, v^{(\sigma)}$, $T^{(\sigma)}_{\rho\mu\nu}$ and $D_{(\mu} \sigma_{\nu)\rho}$ if all of the eigenvalues of $M_{IJ}$ and the factor $\left( 1 - 96 \gamma/\ell^2\right)$ are non-negative. The eigenvalues of the matrix $M_{IJ}$ are explicitly given by
\begin{align}
 \lambda_{\pm} = \left(\frac{3 d - 7}{4} + 48 (d-4)(d-3) \frac{\gamma}{\ell^2} \pm \sqrt{ \frac{(d-1)^2}{16} + 48^2 (d-4)^2 (d-3)^2 \frac{\gamma^2}{\ell^4}} \right).
\end{align}
One can see that $\lambda_{+}$ is always positive and $\lambda_{-}$ is non-negative if and only if
\begin{align}
 \ell^2 \geq - \frac{48(d-4)(3 d - 7)}{d - 2} \gamma.
\end{align} 

To summarize, the quantum focusing conjecture is satisfied for any choice of $v^{(\theta)}, v^{(\sigma)}$, $T^{(\sigma)}_{\rho\mu\nu}$ and $D_{(\mu} \sigma_{\nu)\rho}$, if the smearing scale $\ell$ is large enough to satisfy 
\begin{align}
 \ell^2 \geq 96 \gamma,  \qquad  \ell^2 \geq - \frac{48 (d - 4)(3 d - 7)}{d - 2} \gamma.
\end{align}
Depending on the sign of $\gamma$, either one of the two inequality is trivially satisfied. Thus, the quantum focusing conjecture requires
\begin{align}
 \ell &\geq \ell_{+} := 4\sqrt{6 \gamma}, &(\text{for}~\gamma > 0),\label{lp} \\
 \ell &\geq \ell_{-} := \sqrt{48 \frac{(3d - 7) (d - 4)}{d-2} |\gamma|}, &(\text{for}~\gamma < 0).\label{lm}
\end{align}
This is the main result of this paper. The possible smallest value of the smearing scale, that is $\ell_{+}$ or $\ell_{-}$ depending on the sign of $\gamma$, is understood as the cutoff scale of the higher curvature gravity. We note that $\ell_{-}$ depends on the spacetime dimension $d$. For example, $d = 5$, the cutoff scale for $\gamma < 0$ case is given by $\ell_{-} = 8 \sqrt{2 |\gamma|}$.

\section{Example: Five Dimensional Schwarzschild Spacetime}
\label{example}
\subsection{Setup}
In this section, we investigate an explicit example for evaluation of the derivative of the smeared quantum expansion and see that the smeared quantum expansion satisfies the quantum focusing conjecture if the smearing scale satisfies \eqref{lp} or \eqref{lm}.
We consider the Schwarzschild spacetime as the $\mathcal{O}(\epsilon^0)$ term of the metric. Then, let $\sigma$ be an axially symmetric three dimensional surface in the 
five dimensional Schwarzschild spacetime. 
For this purpose, let us describe the metric in the cylindrical coordinates,
\begin{align}
ds^2=-f(r)dt^2+&\frac{1-f(r)}{f(r)}\frac{(zdz+\rho d\rho)(zdz+\rho d\rho)}{r^2}+dz^2+d\rho^2+\rho^2(d\phi^2+\sin^2 \phi~ d\psi^2),
\end{align}
where $f(r)$ is the function of $r$ given by
\begin{align}
f(r)&:= 1-\frac{2 G M}{r^2},
\end{align}
and $r$ is described by $z$ and $\rho$ by
\begin{align}
r := \sqrt{z^2+\rho^2}.
\end{align}
Note that the Schwarzschild coordinates $\{r, \theta\}$ can be obtained by $z = r \cos \theta$ and $\rho = r \sin \theta$.
In the following calculation, we simplify the expression by using $f'(r) = 2 ( 1 - f(r) )/ r$.

 Let us consider the spacelike hypersurface $\Sigma$ defined by $t=t_0$ and the timelike hypersurface $S$ defined by
\begin{align}
z = z(\rho) 
\end{align}
We define the co-dimension $2$ surface $\sigma$ as the cross section of $S$ by $\Sigma$. We will determine the functional form of $z(\rho)$ later.

Let us construct a null hyper surface $N$ as is the previous sections. To express the null vector field on $\sigma$, let us introduce the unit normal vector fields with respect to $\Sigma$ and $S$ by
\begin{align}
t_{\mu} dx^{\mu} &:= - \sqrt{f(r)} dt, \\
n_{\mu} dx^{\mu} &:= n_{z}(z, \rho) d(z - z(\rho)) = n_{z}(z, \rho) (d z - z' d \rho), 
\end{align}
where $n_{z}$ is given by
\begin{align}
n_{z}(z, \rho) = \frac{r}{\sqrt{(\rho+z z'(\rho))^2+f(r) (z-\rho z'(\rho))^2}}.
\end{align}
Note that the definition of $t_{\mu}$ and $n_{\mu}$ can be applied beyond the surface $\Sigma$ or $S$. In such a case, $t_{\mu}$ and $n_{\mu}$ are the unit normal vector of $t = \text{constant}$ slices and $z - z(\rho) = \text{constant}$ slices respectively.
$t_{\mu}$ and $n_{\mu}$ satisfy $t_{\mu} t^{\mu} = -1, n_{\mu} n^{\mu} = 1$ and $t_{\mu} n^{\mu} = 0$ by the definition.
Then, we introduce the outgoing null vector $k_{\mu}$ and the ingoing null vector $l_{\mu}$ on $\sigma$ by
\begin{align}
k^{\mu}|_{\sigma} &:= \left. \frac{\alpha}{\sqrt{2}} (t^{\mu} + n^{\mu}) \right|_{\sigma} ,\label{kSsigma} \\
l^{\mu}|_{\sigma}  &:= \left. \frac{1}{\sqrt{2} \alpha}(t^{\mu} - n^{\mu}) \right|_{\sigma},
\end{align}
which satisfy $k_{\mu} k^{\mu} = 0, l_{\mu} l^{\mu} = 0$ and $k_{\mu}l^{\mu} = - 1$ on $\sigma$.
Here $\alpha$ represents a degree of freedom for the choice of null vector $k^{\mu}$. We can freely choose the coefficient $\alpha$. In the following, we use
\begin{align}
 \alpha = \frac{\sqrt{2}}{n_{z}(z, \rho)},
\end{align}
for the simplicity to express the result.
The tangent vector $k^{\mu}$ of the null generators of $N$, is, then, defined by solving the geodesic equation with the initial condition on $\sigma$ given by \eqref{kSsigma}.

\subsection{Induced Metric}
Let us evaluate the induced metric on $\sigma$.
By the definition, we can express $h_{\mu\nu}$ as the degenerate tensor in the five dimensional spacetime,
\begin{align}
 h_{\mu\nu}dx^{\mu} dx^{\nu} &= (g_{\mu\nu} + k_{\mu} l_{\nu} + l_{\mu} k_{\nu} ) dx^{\mu} dx^{\nu} \notag\\
& =  \frac{\Bigl( \bigl(z (\rho + z z') - \rho f(r) (z - \rho z') \bigr)dz + \bigl(\rho(\rho + z z') + z f(r) (z - \rho z')\bigr) d\rho \Bigr)^2}{r^2 f(r) \Bigl( (\rho + z z')^2 + f(r) (z - \rho z')^2 \Bigr)} \notag\\
& \qquad + \rho^2 (d \phi^2 + \sin^2 \phi~d\psi^2). 
\end{align}
On the other hand, it is also useful to evaluate the expression of the induced metric as the non-degenerate tensor in the three-dimensional space $\sigma$, through the pull back by the embedding $\{y^{i} \} = (\rho, \phi, \psi) \mapsto \{x^{\mu} \} = (t_{0}, z(\rho), \rho, \phi, \psi)$.
The result is
\begin{align}
 h_{ij} dy^{i} dy^{j} = \frac{(\rho + z z')^2 + f(r) (z - \rho z' )^2}{r^2 f(r)}  d\rho^2 + \rho^2 (d \phi^2 + \sin^2 \phi~d \psi^2).\label{hijSch}
\end{align}
We use the same notation $h$ for both the five-dimensional tensor $h_{\mu\nu}$ and its pull back $h_{ij}$.

\subsection{Second Fundamental Form}
To apply the discussion in the previous section, we require that there is a point $p$ on $\sigma$, denoting the coordinate value $\rho = \bar{\rho}$, where the second fundamental form along $k$ direction vanishes \eqref{extrinsic curvature}.
We will determine the functional form of $z(\rho)$ near the point $p$, that is, the coefficients of the Taylor series, 
\begin{align}
 z(\rho) = \bar{z} + z'(\bar{\rho}) (\rho - \bar{\rho}) + \frac{1}{2} z''(\bar{\rho}) (\rho - \bar{\rho})^2 + \frac{1}{3 !} z'''(\bar{\rho}) (\rho - \bar{\rho})^3 + \cdots 
\end{align}
 by this requirement.

To evaluate the second fundamental form on $\sigma$, we will use the null vector field $u^{\mu}$ defined by
\begin{align}
 u^{\mu} := \frac{\alpha}{\sqrt{2}} \left( t^{\mu} + n^{\mu} \right),
\end{align}
instead of $k^{\mu}$. Since $u^{\mu} |_{\sigma} = k^{\mu} |_{\sigma}$,
we obtain
\begin{align}
 K_{\mu\nu}^{(k)}|_{\sigma} = K_{\mu\nu}^{(u)} |_{\sigma} = h_{\mu}{}^{\rho} h_{\nu}{}^{\sigma} \nabla_{\rho} u_{\sigma} |_{\sigma}.
\end{align}
Then, by using the expression of $u^{\mu}$, the pull-back of $K_{\mu\nu}^{(k)}$ to $\sigma$, denote $K_{ij}^{(k)}$, can be evaluated as 
\begin{align}
K^{(k)}_{ij} dy^{i} dy^{j} = \hat{K}^{(k)}_{\rho\rho}~h_{\rho\rho} d\rho^2 + \hat{K}^{(k)}_{\phi\phi} ~ \left( h_{\phi\phi} d\phi^2 + h_{\psi \psi} d\psi^2 \right).
\end{align}
Here, $\hat{K}^{(k)}_{\rho\rho}$ and $\hat{K}^{(k)}_{\phi\phi}$ are the components of $K^{(k)}_{ij}$ based on the tetrad basis, which are given by 
\begin{align}
 \hat{K}^{(k)}_{\rho\rho} &= \frac{(1 - f(r))(z - \rho z')((\rho + z z')^2 - f(r) (z - \rho z')^2) - r^4 f(r) z''}{r^2 \left( (\rho + z z')^2 + f(r) (z - \rho z')^2 \right)} ,\\
 \hat{K}^{(k)}_{\phi\phi} &= - \frac{z(\rho+z z')-f(r) \rho(z - \rho z')}{r^2 \rho},
\end{align}
while $h_{\rho\rho}$, $h_{\phi\phi}$ and $h_{\psi\psi}$ are the components of $h_{ij}$ given by Eq.~\eqref{hijSch}.
In the above expression, $z$ and $r$ are regarded as a function of $\rho$: $z = z(\rho)$ and $r = r (\rho) = \sqrt{z(\rho)^2 + \rho^2}$.  
Thus the requirement $K_{ij}^{(k)}|_{p} = 0$ can be achieved by choosing the functional form of $z(\rho)$ so that
\begin{align}
z'(\bar{\rho}) &= - \frac{\bar{z} \bar{\rho} (1 - f(\bar{r}))}{\bar{z}^2 + f(\bar{r}) \bar{\rho}^2},\label{zdash} \\
z''(\bar{\rho}) &= - \frac{\bar{r}^2 \bar{z}(1 - f(\bar{r}))(\bar{z}^2 - \bar{\rho}^2 f(\bar{r}))}{(\bar{z}^2+f(\bar{r}) \bar{\rho}^2)^3},\label{zddash}
\end{align}
where we use the notion $\bar{r} = \sqrt{\bar{z}^2 + \bar{\rho}^2}$. Finally, the expansion and the shear can be written as 
\begin{align}
 \theta &= \hat{K}^{(k)}_{\rho\rho} + 2 \hat{K}^{(k)}_{\phi\phi}, 
\end{align}
\begin{align}
  \sigma_{ij}dy^{i} dy^{j} &= \left(\hat{K}_{\rho\rho}^{(k)} - \hat{K}^{(k)}_{\phi\phi}\right) \left( \frac{2}{3} h_{\rho\rho} d\rho^2 - \frac{1}{3}h_{\phi\phi} d\phi^2 - \frac{1}{3} h_{\psi \psi} d\psi^2 \right).
\end{align}

\subsection{Derivative of the Second Fundamental Form}
Let us evaluate the derivative of the sear tensor $\sigma_{ij}$ at the point $p$.
Since $\sigma_{ij}$ vanish at the point $p$, we can evaluate the covariant derivative at the point $p$ by the partial derivative,
\begin{align}
 D_{i} \sigma_{jk}|_{p} &= \partial_{i} \sigma_{jk}|_{p}.
\end{align}
Then, we obtain,
\begin{align}
&\left. (D_{i} \sigma_{jk}) dy^{i} \otimes dy^{j} \otimes dy^{k} \right|_{p}  \notag\\
&= \left. \left(\partial_{\rho} \hat{K}^{(k)}_{\rho\rho} - \partial_{\rho} \hat{K}^{(k)}_{\phi\phi} \right)
\left(
\frac{2}{3} h_{zz} d\rho \otimes d \rho \otimes d\rho
- \frac{1}{3} h_{\phi\phi} d\rho \otimes d \phi \otimes d\phi - \frac{1}{3} h_{\psi\psi} d\rho \otimes d \psi \otimes d \psi \right) \right|_{p}.
\end{align}
Similarly, the derivative of the expansion can be evaluated as
\begin{align}
\left.  D_{i} \theta d y^{i} \right|_{p} = \left. \left( \partial_{\rho} \hat{K}_{\rho\rho}^{(k)} + 2 \partial_{\rho} \hat{K}^{(k)}_{\phi\phi} \right) d \rho \right|_{p}.
\end{align}

Let us focus on the case where the contribution from the $D_{(\mu} \sigma_{\nu)\rho}$ in $\langle d \Theta/ d\lambda \rangle|_{p}$ vanishes.
It can be achieved by choosing the third derivative of $z(\rho)$ at the point $p$ so that
\begin{align}
\left. \partial_{\rho} \hat{K}^{(k)}_{\rho\rho} \right|_{p} = \left. \partial_{\rho} \hat{K}^{(k)}_{\phi\phi} \right|_{p}.\label{zthird}
\end{align}
More explicitly, this condition can be expressed as
\begin{align}
 z'''(\bar{\rho}) & = - \frac{3 \bar{z} \bar{\rho} \bar{r}^2 (1 - f(\bar{r}))}{(\bar{z}^2 + f(\bar{r}) \bar{\rho}^2)^5} \Bigl( (1 - 3 f(\bar{r})) \bar{z}^4 - 2 ( 1 - f(\bar{r})) f(\bar{r}) \bar{z}^2 \bar{\rho}^2 + f(\bar{r})^2 (1 + f(\bar{r})) \bar{\rho}^4  \Bigr).\label{zdddash}
\end{align}

Since all of the components of $D_{i}\sigma_{jk}$ vanish under the condition \eqref{zthird}, we obtain
\begin{align}
 v^{(\sigma)}_{i} |_{p} = T^{(\sigma)}_{ijk}|_{p} = D_{(i} \sigma_{j)k}|_{p} = 0.\label{vTsigmaSch}
\end{align}
In addition, the components described by $D_{i} \theta$ can be evaluated as 
\begin{align}
\left.  v^{(\theta)}_{i} dy^{i} \right|_{p}  &= \left. \frac{1}{3} D_{i} \theta dy^{i} \right|_{p}  = \left. \partial_{\rho} \hat{K}^{(k)}_{\phi\phi}  d \rho \right|_{p}
= \frac{2 \bar{z} \bar{\rho} (1 - f(\bar{r})) f(\bar{r})}{(\bar{z}^2 + \bar{\rho}^2 f(\bar{r}))^2}~d\rho
.\label{vthetaSch}
\end{align}

Finally, the independent components of the Weyl tensor, $v_{i}$ and $T_{ijk}$, can be evaluated through the Codazzi equation, 
\begin{align}
 v_{i}|_{p} = v^{(\theta)}_{i}|_{p},\qquad T_{ijk}|_{p} = 0.\label{vT5dSch}
\end{align}

\subsection{Derivative of the Quantum Expansion}
Substituting Eqs.~\eqref{vT5dSch} into Eq.~\eqref{the derivative QE}, the first derivative of quantum expansion at the point $p$ can be evaluated as
\begin{align}
 \left. \frac{d \Theta}{ d\lambda} \right|_p &= - 8 \gamma v_{i} v^{i}|_{p}
= - 32 \gamma \frac{\bar{z}^2 \bar{\rho}^2 (1 - f(\bar{r}))^2 f(\bar{r})^2}{r^2 (\bar{z}^2 + f(\bar{r}) \bar{\rho}^2)^3}.
 \end{align}
Thus, our surface $\sigma$ with the conditions~\eqref{zdash}, \eqref{zddash}, and \eqref{zdddash} is an example of the violation of the quantum focusing conjecture for unsmeared quantum expansion at the point $p$, if $\gamma$ is negative.

By substituting Eqs~\eqref{vTsigmaSch} and \eqref{vthetaSch} into Eq.~\eqref{smeared Q}, the derivative of the smeared quantum expansion can be obtained as
\begin{align}
 \left \langle \frac{d \Theta}{d \lambda} \right\rangle_{\ell} = - (\ell^2 + 32 \gamma) \frac{\bar{\rho}^2 \bar{z}^2 (1 - f(\bar{r}))^2 f(\bar{r})^2}{\bar{r}^2 (\bar{z}^2 + f(\bar{r}) \bar{\rho}^2)^3}.
\end{align}
Thus the quantum focusing conjecture is satisfied even when $\gamma < 0$ if the smearing scale $\ell$ satisfies the inequality,
\begin{align}
\ell \geq 4 \sqrt{2 |\gamma|}.
\end{align}
We note that this is automatically satisfied if $\ell$ satisfies our requirement \eqref{lm} for $d = 5$, that is, $\ell \geq \ell_{-} = 8 \sqrt{2 |\gamma|}$.

To summarize, if the coupling $\gamma$ is negative, our axially symmetric surface $\sigma$ in the five dimensional Schwarzschild spacetime, with the conditions \eqref{zdash}, \eqref{zddash}, and \eqref{zdddash}, provides an example where the quantum focusing conjecture does not hold for unsmeared quantum expansion but holds for smeared over the length satisfies Eq.~\eqref{lm}.

\section{Summary and Discussion}
\label{summary}
In this paper, we derive the condition for the averaging scale $\ell$ by demanding that the quantum focusing conjecture holds for the quantum expansion smeared over $\ell$ in the quadratic gravity in the classical limit $\hbar \rightarrow 0$.
The gravitational action that we focused on is given by Eq.~\eqref{quadratic gravity} and
the corresponding gravitational entropy is given by Eq.~\eqref{entropy of quadratic gravity}. These expression includes three parameters $\alpha$, $\beta$ and $\gamma$.
At the beginning of Sec.~\ref{sec:QE}, we find that the $\alpha$ and $\beta$ terms, and more generally any contribution of the form of Eq.~\eqref{DeltaS}, do not contribute to the quantum expansion and its derivative at the point $p$ where the $K^{(l)}_{\mu\nu}$ vanishes. The expression of the first derivative of the quantum expansion is given by Eq.~\eqref{QexpressedvT} and reduces to Eq.~\eqref{the derivative QE} at the point $p$. The expression~\eqref{the derivative QE} agrees with the results in Refs.~\cite{Fu:2017lps, Leichenauer:2017bmc} for the Gauss--Bonnet gravity, because $\alpha$ and $\beta$ terms are irrelevant, and hence the quantum focusing conjecture could be violated at the point $p$. Then, motivated by the observation in Ref.~\cite{Leichenauer:2017bmc}, we rigorously formulate the smearing procedure by using a $(d-2)$ dimensional hypercube with the side length $\ell$ in the Riemannian normal coordinates, and evaluate the first derivative of the smeared quantum expansion. The final expression is given by Eq.~\eqref{smeared Q}. By demanding that the non-positivity of Eq.~\eqref{smeared Q}, that is nothing but the quantum focusing conjecture, for any choice of the spacetime and the surface, we obtain the sufficient conditions for the smearing scale $\ell$, Eqs.~\eqref{lp} and \eqref{lm}. In the section \ref{example}, we investigate an axially symmetric surface in the five dimensional Schwarzschild spacetime as an example where the quantum focusing conjecture at a point $p$ is violated but the smeared one is satisfied.

One observation from our final result, \eqref{lp} and \eqref{lm}, is that the $d$ dependence of the cutoff scale $\ell_{\pm}$. Let us compare our results with the argument of the cutoff scale required from the view point of causality~\cite{Camanho:2014apa}.
The positivity of the time delay for the graviton passing through a shock wave requires that the impact parameter $b$ must be greater than the scale determined by the coupling constant $\gamma$: $b \geq b_{+} := 2 \sqrt{(d-4) \gamma}$ for $\gamma > 0$ and $b \geq b_{-} := 2 (d-4) \sqrt{|\gamma|}$ for $\gamma < 0$
\footnote{
These inequalities are derived from Eq.~(2.19) in Ref.~\cite{Camanho:2014apa}.
The former inequality is required when $\epsilon_{ij} n^{j} = 0$ and the latter one is  $\frac{\epsilon_{ij} n^{j} \epsilon_{i}{}_{k} n^{k}}{ \epsilon_{lm} \epsilon_{lm}} = \frac{d-3}{d-2}$, where $\epsilon_{ij}$ is the polarization tensor of the graviton, $n^{i} = b^{i}/|b|$ and $b^{i}$ is the impact parameter vector. Here $i,j, \dots $ represent the indices in the Cartesian coordinates of $d - 2$ dimensional spacetime transverse to the graviton propagation. }
.
Comparing the two length scale $l_{\pm}$ and $b_{\pm}$, we can say that the cutoff length scale required by the quantum focusing conjecture is larger than the one required by the causality at least when $d$ is not so large. Concretely speaking, one can see that $l_{+} \geq b_{+}$ for $5 \leq  d \leq 28$ and $l_{-} > b_{-}$ for $5 \leq d \leq 39$.
In this sense, our result implies that the quantum focusing conjecture is stronger than the causality. 

Throughout the paper, we have focused on the $d \geq 5$ dimensional spacetime. Then, it might be interesting to ask what happens for $d = 4$ case. 
As we commented in the main section, the contributions from the quadratic gravity disappear in $d = 4$ for the perturbative solution around the Ricci flat spacetime. Thus, to see the leading order correction, we need to consider the higher derivative corrections at the cubic order in the curvature tensor or consider matter field, e.g. electromagnetic field, as a source of the Ricci tensor.
In the quadratic gravity, the Codazzi equation plays an important role to relate the smearing effects, $v^{(\theta)}_{i}, v^{(\sigma)}_{i}$ and $T^{(\sigma)}_{ijk}$, and the components of the Weyl tensor, $v$ and $T_{ijk}$, describing the violation of the quantum focusing conjecture at the point $p$. 
It is not clear whether similar relation appears even for the cubic order gravity or even when the matter fields are included. We do not address this issue in the present paper, leaving it for the future work.

\begin{acknowledgments}
T.K. and D.Y. would like to thank Tetsuya Shiromizu and Keisuke Izumi for helpful comments on this work.
T.K. is supported by Nagoya University Interdisciplinary Frontier Fellowship implemented by JST and Nagoya University.
This work is supported by JSPS KAKENHI Grant Numbers JP20K03975 (K.M.), JP20H01902 (T.N.), JP22H01220 (T.N.) and JP20K14469 (D.Y.), MEXT KAKENHI Grant Numbers JP21H05186 (K.M.), JP21H05184 (T.N.), JP23H04007 (T.N.) and JP21H05189 (D.Y.) and Japan--India Cooperative Scientific Programme between JSPS and DST Grant Number JPJSBP120227705 (T.N and D.Y.).
\end{acknowledgments}

\appendix

\section{Some Formulae on $2 + (d-2)$ Decomposition}
\label{sec:A}
In this appendix, we derive the formulae about the $2 + (d-2)$ decomposition
used in the body of this paper.
The formulae in this appendix can be derived in the following setup:
\\
1. $k^{\mu}$ is the tangent of affine parametrized null geodesics.
Thus, we assume that $k^{\mu}$ satisfies
\begin{align}
 k^{\mu} k_{\mu} = 0, \qquad  k^{\nu} \nabla_{\nu} k^{\mu} = 0,
\end{align}
\\
2. $l^{\mu}$ is a null vector field which satisfies
\begin{align}
 l^{\mu} l_{\mu} = 0, \qquad k^{\mu} l_{\mu} = -1.
\end{align}
3.
$k^{\mu}$ and $l^{\mu}$ are orthogonal to a co-dimension 2 surface with the induced metric $h_{\mu\nu}$.
Then, the induced metric $h_{\mu\nu}$ is given by 
\begin{align}
 h_{\mu\nu} = g_{\mu\nu} + k_{\mu} l_{\nu} + l_{\mu} k_{\nu},
\end{align}
$k^{\mu}$ and $l^{\mu}$ are orthogonal to $h_{\mu\nu}$, $k^{\mu} h_{\mu\nu} = l^{\mu} h_{\mu\nu} = 0$.
In addition, $k_{\mu}$ and $l_{\mu}$ are twist-free:
\begin{align}
 K^{(k)}_{[\mu\nu]} = K^{(l)}_{[\mu\nu]} = 0,
\end{align}
where the second fundamental forms are defined as in the main section, see Eq.~\eqref{defKkKl}.

\subsection{Lie Derivative of $h_{\mu\nu}$}
Let us evaluate the Lie derivative of $h_{\mu\nu}$ projected to $\sigma$.
It can be calculated as
\begin{align}
h_{\mu}{}^{\rho} h_{\nu}{}^{\sigma} \mathsterling_{k} h_{\rho\sigma} &=
h_{\mu}{}^{\rho} h_{\nu}{}^{\sigma}
\left( 
k^{\lambda} \nabla_{\lambda} h_{\rho\sigma} + h_{\lambda\sigma} \nabla_{\rho} k^{\lambda} + h_{\rho\lambda} \nabla_{\sigma} k^{\lambda}
\right)  \\
&= 0 +  K_{\mu\nu}^{(k)} + K_{\nu\mu}^{(k)} = 2 K^{(k)}_{\mu\nu},
\end{align}
where we use the fact that 
$h_{\mu}{}^{\rho} h_{\nu}{}^{\sigma} \nabla_{\lambda} h_{\rho\sigma} = h_{\mu}{}^{\rho} h_{\nu}{}^{\sigma} \nabla_{\lambda} (k_{\rho} l_{\sigma} + l_{\rho} k_{\sigma}) = 0$ and $K^{(k)}_{\mu\nu}$ is symmetric.
Thus we obtain
\begin{align}
 h_{\mu}{}^{\rho} h_{\nu}{}^{\sigma} \mathsterling_{k} h_{\rho\sigma} &= 2 K^{(k)}_{\mu\nu}.\label{Lieh}
\end{align}
By similar way, we can also obtain
\begin{align}
 h^{\mu}{}_{\rho} h^{\nu}{}_{\sigma} \mathsterling_{k} h^{\rho\sigma} = - 2K^{(k)}{}^{\mu\nu}.\label{Liehup}
\end{align}

\subsection{Lie Derivative of $K$}
\begin{align}
 h_{\mu}{}^{\rho}h_{\nu}{}^{\sigma} \mathsterling_{k} K^{(k)}_{\rho\sigma}
&= h_{\mu}{}^{\rho}h_{\nu}{}^{\sigma} \left( k^{\lambda} \nabla_{\lambda} K^{(k)}_{\rho\sigma}
+ K^{(k)}_{\lambda \sigma} \nabla_{\mu} k^{\lambda}
+ K^{(k)}_{\mu \lambda} \nabla_{\sigma} k^{\lambda}
 \right) \notag\\
&= h_{\mu}{}^{\rho}h_{\nu}{}^{\sigma}  k^{\lambda} \nabla_{\lambda} K^{(k)}_{\rho\sigma} + 2 K_{\mu\rho}^{(k)} K^{(k)}_{\nu}{}^{\rho}. 
\end{align}
The first term in the last line can be calculated as 
\begin{align}
 h_{\mu}{}^{\rho}h_{\nu}{}^{\sigma}  k^{\lambda} \nabla_{\lambda} K^{(k)}_{\rho\sigma} & = 
 h_{\mu}{}^{\rho}h_{\nu}{}^{\sigma}  k^{\lambda} \nabla_{\lambda} \left( h_{\rho}{}^{\alpha} h_{\sigma}{}^{\beta} \nabla_{\alpha} k_{\beta} \right) \notag\\
& =  h_{\mu}{}^{\alpha}h_{\nu}{}^{\beta}  k^{\lambda}  \nabla_{\lambda}  \nabla_{\alpha} k_{\beta}, \notag\\
& =  h_{\mu}{}^{\alpha}h_{\nu}{}^{\beta}  k^{\lambda}  R_{\lambda\alpha\beta}{}^{\gamma} k_{\gamma} + h_{\mu}{}^{\alpha}h_{\nu}{}^{\beta}  k^{\lambda}  \nabla_{\alpha}  \nabla_{\lambda} k_{\beta},
\end{align}
where we use the affine parametrized geodesic equation $k^{\nu} \nabla_{\nu} k^{\mu} = 0$ and $k^{\nu} \nabla_{\mu} k_{\nu} = 0$ in the second equality.
Further more, the last term can be evaluated as
 \begin{align}
 h_{\mu}{}^{\alpha}h_{\nu}{}^{\beta}  k^{\lambda}  \nabla_{\alpha}  \nabla_{\lambda} k_{\beta} &= -  h_{\mu}{}^{\alpha}h_{\nu}{}^{\beta}  \nabla_{\alpha} k^{\lambda}    \nabla_{\lambda} k_{\beta} \notag\\
&= -  h_{\mu}{}^{\alpha}h_{\nu}{}^{\beta} \left( h_{\rho}{}^{\lambda} - k_{\rho} l^{\lambda} - l_{\rho} k^{\lambda} \right) \nabla_{\alpha} k^{\rho}    \nabla_{\lambda} k_{\beta} \notag\\
 &= - K^{(k)}_{\mu}{}^{\rho} K^{(k)}_{\rho\nu}.
 \end{align}
Thus, we obtain the useful formula, 
\begin{align}
 h_{\mu}{}^{\rho}h_{\nu}{}^{\sigma} \mathsterling_{k} K^{(k)}_{\rho\sigma}
= h_{\mu}{}^{\alpha}h_{\nu}{}^{\beta}  k^{\lambda}  R_{\lambda\alpha\beta}{}^{\gamma} k_{\gamma} + K_{\mu\rho}^{(k)} K^{(k)}_{\nu}{}^{\rho}. \label{LieKmunu}
\end{align}

By using Eq.~\eqref{LieKmunu}, as well as Eq.~\eqref{Liehup}, we obtain
\begin{align}
\mathsterling_{k} K^{(k)} & =
 h^{\mu\nu} \mathsterling_{k} K^{(k)}_{\mu\nu} + \mathsterling_{k} h^{\mu\nu} K^{(k)}_{\mu\nu} \notag\\
&= 
h^{\alpha\beta} R_{\lambda\alpha\beta\tau} k^{\lambda}k^{\tau} + K^{(k)}_{\mu\nu} K^{(k)}{}^{\mu\nu} - 2 K^{(k)}{}^{\mu\nu} K^{(k)}_{\mu\nu} \notag\\
&=
 - K^{(k)}{}^{\mu\nu} K^{(k)}_{\mu\nu} - R_{\mu\nu} k^{\mu} k^{\nu}. \label{LieK}
\end{align}
Note that Eq.~\eqref{LieK} is nothing but the Raychaudhuri equation \eqref{Raychaudhuri} if we express the second fundamental form $K^{(k)}_{\mu\nu}$ by $\theta$ and $\sigma_{\mu\nu}$.

\subsection{Gauss Equations}
By expressing the covariant derivative $D_{\mu}$ with respect to $h_{\mu\nu}$ in terms of the covariant derivative $\nabla_{\mu}$ with respect to $g_{\mu\nu}$, we obtain
\begin{align}
 D_{\mu} D_{\nu} V_{\rho} &= h_{\mu}{}^{\alpha} h_{\nu}{}^{\beta} h_{\rho}{}^{\gamma}  \nabla_{\alpha} \left( h_{\beta}{}^{\lambda} h_{\gamma}{}^{\tau} \nabla_{\lambda} V_{\tau} \right), \\
&= 
h_{\mu}{}^{\alpha} h_{\nu}{}^{\beta} h_{\rho}{}^{\tau}  \nabla_{\alpha}  h_{\beta}{}^{\lambda} \nabla_{\lambda} V_{\tau} 
+
h_{\mu}{}^{\alpha} h_{\nu}{}^{\lambda} h_{\rho}{}^{\gamma} \nabla_{\alpha} h_{\gamma}{}^{\tau} \nabla_{\lambda} V_{\tau} 
+
h_{\mu}{}^{\alpha} h_{\nu}{}^{\lambda} h_{\rho}{}^{\tau}  \nabla_{\alpha} \nabla_{\lambda}  V_{\tau},
\end{align}
where $V_{\mu}$ is an arbitrary vector field on $\sigma$, $V_{\mu} = h_{\mu}{}^{\nu} V_{\nu}$.
By expressing the $h_{\mu}{}^{\nu}$ that the covariant derivative is acting on in terms of $k$ and $l$, we obtain following expression
\begin{align}
 D_{\mu} D_{\nu} V_{\rho} &= K_{\mu\nu}^{(l)} h_{\rho}{}^{\tau} k^{\lambda} \nabla_{\lambda} V_{\tau} + K^{(k)}_{\mu\nu} h_{\rho}{}^{\tau} l^{\lambda} \nabla_{\lambda} V_{\tau} \notag \\
&\qquad  - K^{(k)}_{\mu\rho} K^{(l)}_{\nu}{}^{\sigma} V_{\sigma} 
- K^{(l)}_{\mu\rho} K^{(k)}_{\nu}{}^{\sigma} V_{\sigma}
+
h_{\mu}{}^{\alpha} h_{\nu}{}^{\beta} h_{\rho}{}^{\gamma}  \nabla_{\alpha} \nabla_{\beta}  V_{\gamma}.
\end{align}
Here used the identity $k^{\mu} \nabla_{\nu} V_{\mu} = - \nabla_{\nu} k^{\mu} V_{\mu}$ and so on.

By antisymmetrized the indices $\mu$ and $\nu$, we obtain;
\begin{align}
 {}^{(d-2)}R_{\mu\nu\rho}{}^{\sigma} V_{\sigma}
 &= 2 D_{[\mu}D_{\nu]} V_{\sigma} \notag\\
 & =
2 h_{[\mu}{}^{\alpha} h_{\nu]}{}^{\beta} h_{\rho}{}^{\gamma}  \nabla_{\alpha} \nabla_{\beta}  V_{\gamma}
- 2 K^{(k)}_{[\mu|\rho} K^{(l)}_{|\nu]}{}^{\sigma} V_{\sigma} 
- 2 K^{(l)}_{[\mu|\rho} K^{(k)}_{|\nu]}{}^{\sigma} V_{\sigma} \notag\\
&= \left( h_{\mu}{}^{\alpha} h_{\nu}{}^{\beta} h_{\rho}{}^{\gamma} h_{\delta}{}^{\sigma} R_{\alpha\beta\gamma}{}^{\delta}  - 2 K^{(k)}_{[\mu|\rho} K^{(l)}_{|\nu]}{}^{\sigma}- 2 K^{(l)}_{[\mu|\rho} K^{(k)}_{|\nu]}{}^{\sigma} \right) V_{\sigma}.
\end{align}
In the second equality, we used the fact that $K^{(l)}_{[\mu\nu]} = 0$ as well as $K^{(k)}_{[\mu\nu]} = 0.$ 
Since this equation holds for any vector $V_{\mu}$ on $\sigma$, we obtain the Gauss equation,
\begin{align}
 {}^{(d-2)}R_{\mu\nu\rho}{}^{\sigma} = h_{\mu}{}^{\alpha} h_{\nu}{}^{\beta} h_{\rho}{}^{\gamma} h_{\delta}{}^{\sigma} R_{\alpha\beta\gamma}{}^{\delta}  - 2 K^{(k)}_{[\mu|\rho} K^{(l)}_{|\nu]}{}^{\sigma}- 2 K^{(l)}_{[\mu|\rho} K^{(k)}_{|\nu]}{}^{\sigma}. \label{GaussEquation}
\end{align}
By contracting indices $\nu$ and $\sigma$, we obtain
\begin{align}
 {}^{(d-2)} R_{\mu\rho}&=h_{\mu}{}^{\alpha} h_{\rho}{}^{\gamma} R_{\alpha\gamma}
+ h_{\mu}{}^{\alpha} h_{\rho}{}^{\gamma} R_{\alpha\beta\gamma\delta} l^{\beta} k^{ \gamma}
+ h_{\mu}{}^{\alpha} h_{\rho}{}^{\gamma} R_{\alpha\beta\gamma\delta} k^{\beta} l^{ \gamma} \notag\\
& \qquad - K^{(k)}_{\mu\rho} K^{(l)}
+ K^{(k)}_{\nu\rho} K^{(l)}_{\mu}{}^{\nu}
- K^{(l)}_{\mu\rho} K^{(k)}
+ K^{(l)}_{\nu\rho} K^{(k)}_{\mu}{}^{\nu}.
\end{align}
Again, by contracting the indices $\mu$ and $\rho$, we obtain,
\begin{align}
 {}^{(d-2)} R &= R + 4 k^{\alpha} l^{\gamma} R_{\alpha\gamma} - 2 k^{\alpha} l^{\beta}  k^{ \gamma} l^{\delta} R_{\alpha\beta\gamma\delta} - 2 K^{(k)} K^{(l)} + 2 K^{(k)}_{\mu\nu} K^{(l)}{}^{\mu\nu}.
\end{align}
Note that by using $q_{\mu\nu} = - k_{\mu} l_{\nu} - l_{\mu} k_{\nu}$, above equation can be expressed as 
\begin{align}
 {}^{(d-2)} R = R - 2 q^{\alpha\gamma} R_{\alpha\gamma} + q^{\alpha\gamma} q^{\beta \delta} R_{\alpha\beta\gamma\delta} + K_{\rho} K^{\rho} - K_{\rho\mu\nu} K^{\rho\mu\nu}. \label{GausseqRS}
\end{align}

\subsection{Codazzi Equations}
Similar to the derivation of the Gauss equation we can derive the Codazzi Equation by rewriting the covariant derivative $D_{\mu}$ in terms of $\nabla_{\mu}$.
\begin{align}
 D_{\mu} K^{(k)}_{\nu\rho} &= h_{\mu}{}^{\alpha} h_{\nu}{}^{\beta} h_{\rho}{}^{\gamma} \nabla_{\alpha} \left(h_{\beta}{}^{\lambda}h_{\gamma}{}^{\tau} \nabla_{\lambda} k_{\tau} \right) \notag\\
&= 
h_{\mu}{}^{\alpha} h_{\nu}{}^{\beta} h_{\rho}{}^{\tau}  \nabla_{\alpha} h_{\beta}{}^{\lambda} \nabla_{\lambda} k_{\tau}
+
h_{\mu}{}^{\alpha} h_{\nu}{}^{\lambda} h_{\rho}{}^{\gamma}  \nabla_{\alpha} h_{\gamma}{}^{\tau} \nabla_{\lambda} k_{\tau}
+
h_{\mu}{}^{\alpha} h_{\nu}{}^{\lambda} h_{\rho}{}^{\tau} \nabla_{\alpha} \nabla_{\lambda} k_{\tau}
\end{align}
Again, by expressing the $h_{\mu}{}^{\nu}$ with the covariant derivative by $k$ and $l$, we obtain following expression
\begin{align}
 D_{\mu} K^{(k)}_{\nu\rho} &= K_{\mu\nu}^{(k)} h_{\rho}{}^{\tau} l^{\lambda} \nabla_{\lambda} k_{\tau}
+ K_{\mu\nu}^{(l)} h_{\rho}{}^{\tau} k^{\lambda} \nabla_{\lambda} k_{\tau} \notag\\
&\qquad + K^{(k)}_{\mu\rho} h_{\nu}{}^{\lambda} l^{\tau} \nabla_{\lambda} k_{\tau} +
h_{\mu}{}^{\alpha} h_{\nu}{}^{\beta} h_{\rho}{}^{\gamma} \nabla_{\alpha} \nabla_{\beta} k_{\gamma}.
\end{align}
Then, antisymmetrized the first two indices, we obtain the Codazzi equation
\begin{align}
 D_{[\mu} K^{(k)}_{\nu]\rho} &= h_{[\mu}{}^{\alpha} h_{\nu]}{}^{\beta} h_{\rho}{}^{\gamma} \nabla_{\alpha} \nabla_{\beta} k_{\gamma} + K^{(k)}_{[\mu|\rho} h_{|\nu]}{}^{\lambda} l^{\tau} \nabla_{\lambda} k_{\tau} \notag\\
&= \frac{1}{2} h_{\mu}{}^{\alpha} h_{\nu}{}^{\beta} h_{\rho}{}^{\gamma} R_{\alpha\beta\gamma}{}^{\delta} k_{\delta} + K^{(k)}_{[\mu|\rho} h_{|\nu]}{}^{\lambda} l^{\tau} \nabla_{\lambda} k_{\tau}.
\end{align}

For our purpose, it is useful to express the Codazzi equation in terms of the Weyl tensor and others,
\begin{align}
 D_{[\mu} K^{(k)}_{\nu]\rho} &= \frac{1}{2} \mathcal{C}_{\boldsymbol{k}\rho\nu\mu} + \mathcal{O}(R_{\mu\nu}, K^{(k)}_{\mu\nu}).\label{CodazziWeyl}
\end{align}
Expressing $K^{(k)}_{\mu\nu}$ in terms of the expansion $\theta$ and shear $\sigma_{\mu\nu}$ by Eq.~\eqref{defthetasigma} and rewriting their derivative in terms of $v^{(\theta)}_{\mu}, v^{(\sigma)}_{\mu}$ and $T^{(\sigma)}_{\mu\nu\rho}$ defined by \eqref{vtheta} - \eqref{Tsigma}, the left hand side of the Codazzi equation \eqref{CodazziWeyl} can be expressed as
\begin{align}
D_{[\mu} K^{(k)}_{\nu]\rho} &=  \left(v^{(\theta)}_{[\mu} + v^{(\sigma)}_{[\mu} \right) h_{\nu] \rho}
- \frac{1}{2} T^{(\sigma)}_{\rho\mu\nu}.\label{CodazziLHS}
\end{align}
On the other hand, the right hand of the Codazzi equation \eqref{CodazziWeyl} can be expressed in terms of $v_{\mu}$ and $T_{\mu\nu\rho}$ through Eq.~\eqref{Ck=v+T}, 
\begin{align}
\frac{1}{2} \mathcal{C}_{\boldsymbol{k}\rho\nu\mu} = v_{[\mu} h_{\nu]\rho} - \frac{1}{2} T_{\rho\mu\nu} + \mathcal{O}(R_{\mu\nu}, K^{(k)}_{\mu\nu}).\label{CodazziRHS}
\end{align}
Comparing Eqs.~\eqref{CodazziLHS} and \eqref{CodazziRHS}, we obtain the result,
\begin{align}
 v_{\mu} &= v_{\mu}^{(\theta)} + v^{(\sigma)}_{\mu} + \mathcal{O}(R_{\mu\nu}, K^{(k)}_{\mu\nu}), \qquad 
 T_{\rho\mu\nu} = T^{(\sigma)}_{\rho\mu\nu} +\mathcal{O}(R_{\mu\nu}, K^{(k)}_{\mu\nu}).\label{Codazzi}
\end{align}

\subsection{Other Formulae}
In this subsection, we derive the other formulae used in the main section.

First, let us derive the expression for $C_{\mu\rho\sigma\tau} C_{\nu}{}^{\rho\sigma\tau} k^{\mu} k^{\nu}$ in terms of the projected Weyl tensors $\mathcal{C}_{\boldsymbol{k}\mu\nu\rho}$ and so on.
It can be obtained as follows:
\begin{align}
C_{\mu\rho\sigma\tau} C_{\nu}{}^{\rho\sigma\tau} k^{\mu} k^{\nu}
&=  C_{\mu\rho\sigma\tau} C_{\nu\alpha\beta\gamma} k^{\mu} k^{\nu} g^{\rho\alpha} g^{\sigma\beta} g^{\tau\gamma}
\\
&=  C_{\mu\rho\sigma\tau} C_{\nu\alpha\beta\gamma} k^{\mu} k^{\nu} h^{\rho\alpha} (h^{\sigma\beta} - k^{\sigma} l^{\beta} - l^{\sigma} k^{\beta}) (h^{\tau\gamma} - k^{\tau}l^{\gamma} - l^{\tau} k^{\beta})
\\ 
&= \mathcal{C}_{\boldsymbol{k}\mu\nu\rho} \mathcal{C}_{\boldsymbol{k}}{}^{\mu\nu\rho}
- 2 \mathcal{C}_{\boldsymbol{k} \boldsymbol{l} \nu \boldsymbol{k}} \mathcal{C}_{\boldsymbol{k} \boldsymbol{l}}{}^{\nu}{} _{\boldsymbol{k}} 
- 4 \mathcal{C}_{\boldsymbol{k} \mu \boldsymbol{k} \nu} \mathcal{C}_{\boldsymbol{k}} {}^{\mu}{}_{\boldsymbol{l}}{}^{\nu}. \label{CkCk1}
\end{align}
In addition, the traceless property of the Weyl tensor can be represented as
\begin{align}
0 &= k^{\mu} h_{\nu}{}^{\rho} C_{\mu \alpha \rho \beta} g^{\alpha\beta} \notag\\
 &= k^{\mu} h_{\nu}{}^{\rho} C_{\mu \alpha \rho \beta} h^{\alpha\beta} -   k^{\mu} h_{\nu}{}^{\rho} C_{\mu \alpha \rho \beta} l^{\alpha} k^{\beta} \notag\\
&= \mathcal{C}_{\boldsymbol{k}}{}^{\alpha}{}_{\nu}{}_{\alpha}
- \mathcal{C}_{\boldsymbol{k} \boldsymbol{l}  \nu \boldsymbol{k}}.
\end{align} 
Hence, the second term in Eq.~\eqref{CkCk1} can be expressed by $\mathcal{C}_{\boldsymbol{k}\mu\nu\rho}$ and we obtain,
\begin{align}
C_{\mu\rho\sigma\tau} C_{\nu}{}^{\rho\sigma\tau} k^{\mu} k^{\nu} =
\mathcal{C}_{\boldsymbol{k}\mu\nu\rho} \mathcal{C}_{\boldsymbol{k}}{}^{\mu\nu\rho} - 2 \mathcal{C}_{\boldsymbol{k}}{}^{\mu}{}_{\nu\mu} \mathcal{C}_{\boldsymbol{k}}{}^{\rho \nu}{}_{\rho}
-4 \mathcal{C}_{\boldsymbol{k} \mu \boldsymbol{k} \nu} \mathcal{C}_{\boldsymbol{k}}{}^{\mu}{}_{\boldsymbol{l}}{}^{\nu}. \label{CkCk}
\end{align}

Next, let us evaluate ${}^{(d-2)} G^{\mu\nu} R_{\rho\mu\sigma\nu} k^{\rho} k^{\sigma}$. By the definition of the Einstein tensor, this can be expressed as 
\begin{align}
{}^{(d-2)} G^{\mu\nu} R_{\rho\mu\sigma\nu} k^{\rho} k^{\sigma} 
&= {}^{(d-2)} R_{\alpha\beta\gamma\delta} \left(h^{\alpha\gamma} h^{\beta\mu} h^{\delta\nu} - \frac{1}{2} h^{\mu\nu} h^{\alpha\gamma} h^{\beta\delta} \right) R_{\rho\mu\sigma\nu} k^{\rho} k^{\sigma} \label{GRkk1}
\end{align}
Then, by using the Gauss equation, $(d-2)$-dimensional Riemann tensor can be written by the $d$-dimensional Riemann tensor projected on $\sigma$, with the terms including $K^{(k)}_{\mu\nu}$. Thus, the right hand side of \eqref{GRkk1} can be expressed as 
\begin{align}
R_{\alpha\beta\gamma\delta} \left(h^{\alpha\gamma} h^{\beta\mu} h^{\delta\nu} - \frac{1}{2} h^{\mu\nu} h^{\alpha\gamma} h^{\beta\delta} \right) R_{\rho\mu\sigma\nu} k^{\rho} k^{\sigma} + {\cal O}(K^{(k)}_{\mu\nu}).
\end{align}
In addition, by extracting Weyl tensor term, we obtain
\begin{align}
 {}^{(d-2)} G^{\mu\nu} R_{\rho\mu\sigma\nu} k^{\rho} k^{\sigma} 
&= 
C_{\alpha\beta\gamma\delta} \left(h^{\alpha\gamma} h^{\beta\mu} h^{\delta\nu} - \frac{1}{2} h^{\mu\nu} h^{\alpha\gamma} h^{\beta\delta} \right) C_{\rho\mu\sigma\nu} k^{\rho} k^{\sigma} + {\cal O}(K^{(k)}_{\mu\nu}, R_{\mu\nu}) \\
&= \mathcal{C}_{\alpha}{}^{\mu\alpha\nu} \mathcal{C}_{\boldsymbol{k} \mu \boldsymbol{k} \nu} - \frac{1}{2} \mathcal{C}_{\alpha\beta}{}^{\alpha\beta} \mathcal{C}_{\boldsymbol{k}\mu \boldsymbol{k}}{}^{\mu} + {\cal O}(K^{(k)}_{\mu\nu}, R_{\mu\nu}).
\end{align}
Again, the traceless property of the Weyl tensor can be expressed as 
\begin{align}
 \mathcal{C}_{\alpha}{}^{\mu\alpha\nu} =  \mathcal{C}_{\boldsymbol{k}}{}^{\mu}{}_{\boldsymbol{l}}{}^{\nu} + \mathcal{C}_{\boldsymbol{l}}{}^{\mu}{}_{\boldsymbol{k}}{}^{\nu}, \qquad \mathcal{C}_{\boldsymbol{k} \mu \boldsymbol{k}}{}^{\mu} = \mathcal{C}_{\boldsymbol{k} \boldsymbol{k} \boldsymbol{k} \boldsymbol{l}} +  \mathcal{C}_{\boldsymbol{k} \boldsymbol{l} \boldsymbol{k} \boldsymbol{k}} = 0,
\end{align}
and hence we obtain
\begin{align}
  {}^{(d-2)} G^{\mu\nu} R_{\rho\mu\sigma\nu} k^{\rho} k^{\sigma} 
= 2 \mathcal{C}_{\boldsymbol{k}}{}^{\mu}{}_{\boldsymbol{l}}{}^{\nu} \mathcal{C}_{\boldsymbol{k} \mu \boldsymbol{k} \nu}  + {\cal O}(K^{(k)}_{\mu\nu}, R_{\mu\nu}). \label{GRkk}
\end{align}

\section{Functional Derivative and Lie Derivative}
\label{sec:functionalderivative}
The variation of a functional $S$ of $V$ with respect to $\delta V$ is defined by 
\begin{align}
 \delta S[V; \delta V] := \lim_{\epsilon \rightarrow 0} \frac{S[V + \epsilon\,\delta V] - S[V] }{\epsilon}.
\end{align}
Then, the functional derivative of $S$ by $V(y_{1})$ is defined by
\begin{align}
 \frac{\delta S[V]}{\delta V(y_{1})} := \delta S [V; \delta_{y_{1}}],
\end{align}  
where $\delta_{y_{1}}$ is Dirac's delta function $\delta_{y_{1}} (y) := \delta^{(d-2)}(y - y_{1})$.

Let us focus on the situation where $S[V]$ is represented as an integral of a scalar density constructed by the tensor on $V$ by a covariant manner. By denoting the tensor on $V$ by $\chi_{I}$, our assumption can be expressed as 
\begin{align}
 S[V] = \int_{V} d^{d-2} y\, s(\chi_{I}) .
\end{align}
For our purpose, $\chi_{I} = \{h_{\mu\nu}, K^{(k)}_{\mu\nu}, K^{(l)}_{\mu\nu} \}$ and so on.
Let us consider the foliation of a neighborhood of $V$ in $N$ by the slices $\{ V + \epsilon \delta V \}_{\epsilon}$. We can regards the tensor fields $\chi_{I}$ originally defined on each slices $V + \epsilon \delta V$ as tensor fields on $N$. We can define one parameter family of diffeomorphism on $N$ that maps a point $(\lambda, y)$ to $(\lambda + \epsilon \delta V(y), y)$. Since the surface $V$ and $V + \epsilon \delta V$ are diffeomorphic, the integral over $V + \epsilon \delta V$ can be pull-backed to the integral over $V$,
\begin{align}
 S[V + \epsilon\, \delta V] &= \int_{V + \epsilon\, \delta V} d^{d-2} y \, s(\chi_{I}) = \int_{V} d^{d-2} y \, s(\phi_{\epsilon *} \chi_{I}) \notag\\
& = \int_{V} d^{d-2} y \, s(\chi_{I} + \epsilon \mathsterling_{X} \chi_{I} + \mathcal{O}(\epsilon^2)).
\end{align}
Here $\phi_{\epsilon *}$ is the pull-back with respect to the diffeomorphism $\phi_{\epsilon}$, $\mathsterling_{X}$ is the Lie derivative on $N$, and $X^{\mu}$ is the tangent vector of the integral curve by $\phi_{\epsilon}$.
That is nothing but the $\epsilon$ derivative with fixing $y$ coordinate. Since $\lambda = V(y) + \epsilon \delta V(y)$, we obtain
\begin{align}
 X^{\mu}\partial_{\mu} = \frac{\partial}{\partial \epsilon} =  \frac{\partial \lambda}{\partial \epsilon}  \frac{\partial}{\partial \lambda} = \delta V(y) k^{\mu} \partial_{\mu}.
\end{align}
Note that, since $y$ is constant along null generators, $\epsilon$ is also the affine parameter.

In this view, we can regards $S[V]$ as a functional of $\chi_{I}$. We express it by same name $S$,
\begin{align}
 S[\chi_{I}] := \int_{V} d^{d-2} y \, s(\chi_{I}).
\end{align}
Then the variation of $V$ can be translated into the variation of $\chi_{I}$ as follows;
\begin{align}
 \delta S[V; \delta V] &= \lim_{\epsilon \rightarrow 0} \frac{S[V + \epsilon \delta V] - S[V]}{\epsilon} \\
&= \lim_{\epsilon \rightarrow 0} \frac{S[\chi_{I} + \epsilon \mathsterling_{X} \chi_{I}] - S[\chi_{I}]}{\epsilon} \\
&= \delta S[\chi_{I}; \mathsterling_{X} \chi_{I}].
\end{align}
Thus the variation of the surface $V$ with respect to $\delta V$ can be translated into the variation of the tensor $\chi_{I}$ with respect to $\mathsterling_{X} \chi_{I}$.
Expressing the final term in the functional derivative, we obtain a useful formula,
\begin{align}
 \delta S[V; \delta V] &= \int_{V} d^{(d-2)} y \frac{\delta S[\chi_{I}]}{\delta \chi_{I}(y)} \mathsterling_{X} \chi_{I}(y), \qquad X^{\mu}(y) = k^{\mu}(y)\, \delta V(y).  
\end{align}
By setting $\delta V = \delta_{y_{1}}$, we also obtain 
\begin{align}
 \frac{\delta S[V]}{\delta V(y_{1})} &= \int_{V} d^{(d-2)} y \frac{\delta S[\chi_{I}]}{\delta \chi_{I}(y)} \mathsterling_{X} \chi_{I}(y), \qquad X^{\mu}(y) = k^{\mu}(y)\, \delta^{(d-2)}(y - y_{1}) \label{fdtold}.
\end{align}
In the main section, we use the notion of the functional derivative for the geometrical quantity itself, regarding it as a functional $\chi_{I}(y) = \chi_{I}[V;y]$.
Applying \eqref{fdtold}, the functional derivative can be calculated as 
\begin{align}
 \frac{\delta \chi_{I}(y) }{\delta V(y_{1})} = \mathsterling_{X} \chi_{I} (y), \qquad X^{\mu}(y) = k^{\mu}(y)\, \delta^{(d-2)}(y - y_{1}). 
\end{align}
By using this notion, we simply express Eq.~\eqref{fdtold} as the chain rule, 
\begin{align}
 \frac{\delta S[V]}{\delta V(y_{1})} &= \int_{V} d^{(d-2)} y \frac{\delta S[\chi_{I}]}{\delta \chi_{I}(y)} \frac{\delta \chi_{I}(y) }{\delta V(y_{1})}.
\end{align}
We note that the Lie derivative $\mathsterling_{X}$ is defined in the submanifold $N$ and hence $X^{\mu}$ is defined only on $N$. To calculate $\mathsterling_{X} \chi_{I}$ in terms of $d$-dimensional spacetime, one can simply use any extension of $X^{\mu}$ apart from $N$ and project it to $\sigma$ by $h_{\mu}{}^{\nu}$.

By using above formula, let us evaluate the functional derivative of the area functional $A[V]$ and the $d-2$ dimensional Einstein--Hilbert action.
The functional derivative of the area functional can be calculated as follows,
\begin{align}
 \frac{\delta A[V]}{\delta V(y)} &= \frac{1}{2} \sqrt{h} h^{\mu\nu} \mathsterling_{k} h_{\mu\nu} = \sqrt{h} K^{(k)} = \sqrt{h} \theta,\label{dAdV} \\
\end{align}
where we use Eq.~\eqref{Lieh}.
The functional derivative of the Einstein--Hilbert action can be evaluated as,
\begin{align}
  \frac{\delta I^{\text{EH}}[V]}{\delta V(y)} &= - \sqrt{h} {}^{(d-2)}G^{\mu\nu} \mathsterling_{k} h_{\mu\nu} = - 2 \sqrt{h} {}^{(d-2)}G^{\mu\nu} K^{(k)}_{\mu\nu}, \label{dGdV} 
\end{align}
where we again use Eq.~\eqref{Lieh}.

\bibliography{ref} 
\bibliographystyle{JHEP}

\end{document}